\documentclass[12pt]{article}
\usepackage{axodraw}
\usepackage{epsfig}   
\usepackage{amssymb}

\newcommand{\lsim}{\raisebox{-0.13cm}{~\shortstack{$<$ \\[-0.07cm] $\sim$}}~}
\newcommand{\gsim}{\raisebox{-0.13cm}{~\shortstack{$>$ \\[-0.07cm] $\sim$}}~}
\newcommand{\eqa} {\begin{eqnarray} }
\newcommand{\eqe} {\end{eqnarray}}
\newcommand{\gzw} {G_{i}^{R}}       
\newcommand{\gei} {G_{i}^{L}}       
\newcommand{\geis} {G_{i}^{L\ast}}    
\newcommand{\gzws} {G_{i}^{R\ast}}    
\newcommand{\cei} {C_{1k}^L}           
\newcommand{\czw} {C_{1k}^R}          
\newcommand{\ceis} {C_{1k}^{L\ast}}    
\newcommand{\czws} {C_{1k}^{R\ast}}    
\newcommand{\nei} {N_{l1}^L}         
\newcommand{\nzw} {N_{l1}^R}         
\newcommand{\neis} {N_{l1}^{L\ast}}
\newcommand{\nzws} { N_{l1}^{R\ast}}
\newcommand{\kei} {g_{L}^q}
\newcommand{\kzw} {g_{R}^q}

\newcommand{\kll} {\left(}
\newcommand{\klr} {\right)}
\newcommand{\peke} {p_1\! \cdot\! k_1}
\newcommand{\pepz} {p_1\! \cdot\! p_2}
\newcommand{\pekz} {p_1\! \cdot\! k_2}
\newcommand{\pzke} {p_2\! \cdot\! k_1}
\newcommand{\pzkz} {p_2\! \cdot\! k_2}
\newcommand{\peqe} {p_1\! \cdot\! q_1}

\newcommand{\keqe} {k_1\! \cdot\! q_1}
\newcommand{\kzqe} {k_2\! \cdot\! q_1}
\newcommand{\qeqz} {q_1\! \cdot\! q_2}
\newcommand{\peqz} {p_1\! \cdot\! q_2}
\newcommand{\kzqz} {k_2\! \cdot\! q_2}
\newcommand{\kekz} {k_1\! \cdot\! k_2}
\newcommand{\mtt} {m_{\tilde t_i}}
\newcommand{\nel} {\tilde{\chi}_l^0}
\newcommand{\lsp} {\tilde{\chi}_1^0}
\newcommand{\cpk} {\tilde{\chi}_k^{+}}
\newcommand{\cmk} {\tilde{\chi}_k^{-}}
\newcommand{\mne} {m_{\tilde{\chi}_1^0}}
\newcommand{\mlsp} {m_{\tilde{\chi}_1^0}}
\newcommand{\mnel} {m_{\tilde{\chi}_l^0}}
\newcommand{\mcpk} {m_{\tilde \chi_k^\pm}}
\newcommand{\cth} {\cos\theta}

\begin{document}
\pagestyle{empty}
\begin{flushright}
KIAS--P06008 \\
March 2006
\end{flushright}

\begin{center}
{\large\sc {\bf The Passage of Ultrarelativistic Neutralinos through Matter}}

\vspace{1cm}
{\sc Sascha Bornhauser}$^1$ and {\sc Manuel Drees}$^{1,2}$

\vspace*{5mm}
{}$^1${\it Physikalisches Institut, Universit\"at Bonn, Nussallee 12, D53115
  Bonn,  Germany} \\
\vspace*{3mm}
{}$^2$ {\it Korea Institute of Advanced Studies, School of Physics, Seoul,
  South Korea}
\end{center}

\vspace*{1cm}
\begin{abstract}
  
  The origin of the most energetic cosmic ray events, with $E \gsim 10^{20}$
  eV, remains mysterious. One possibility is that they are produced in the
  decay of very massive, long--lived particles. It has been suggested that
  these so--called ``top--down scenarios'' can be tested by searching for
  ultrarelativistic neutralinos, which would be produced copiously if
  superparticles exist at or near the TeV scale. In this paper we present a
  detailed analysis of the interactions of such neutralinos with ordinary
  matter. To this end we compute several new contributions to the total
  interaction cross section; in particular, the case of higgsino--like
  neutralinos is treated for the first time. We also carefully solve the
  transport equations. We show that a semi--analytical solution that has been
  used in the literature to treat the somewhat analogous propagation of $\tau$
  neutrinos leads to large errors; we therefore use a straightforward
  numerical method to solve these integro--differential equations.

\end{abstract}
\newpage
\setcounter{page}{1}

\pagestyle{plain}
\section{Introduction}

Experiments like Fly's Eye \cite{Bir95} and AGASA \cite{Hay99} have shown the
existence of cosmic rays with energy $E \gsim 10^{20}$ eV, the so--called
ultra high energy (UHE) component of the cosmic ray spectrum. This raises two
questions: how were these UHE particles produced, and how did they manage to
reach Earth?

In order to accelerate ultra--relativistic charged particles to energy $E$,
one needs a magnetic field with strength $B$ extending at least over the
confinement radius $L \propto Z/(B\cdot E)$, where $Z$ is the charge of the
particle \cite{siglrev}. The problem is that very few, if any, objects in the
Universe can be said with confidence to have a sufficiently large $B \cdot L$
to accelerate protons to $10^{20}$ eV, once synchrotron losses in high
$B-$fields have been taken into account.

The second problem is that the arrival directions of UHE events seem to be
distributed more or less homogeneously. Since protons of this energy would not
be deflected much by the magnetic fields in our galaxy, this excludes one or a
few local point sources. A large group of very distant sources, e.g. related
to active galactic nuclei, would indeed yield an essentially homogeneous
distribution. Unfortunately particles produced at such very large distances
should not be able to reach us without losing much of their energy. In
particular, as pointed out in refs.~\cite{gzk}, protons with $E \gsim 5 \cdot
10^{19}$ eV would lose their energy through inelastic scattering on the
photons of the cosmic microwave background (CMB); this is known as GZK effect
or GZK cut--off. The same target also depletes the flux of photons with $E
\gsim 10^{15}$ eV, the most important reaction here being the production of
$e^+e^-$ pairs. Finally, heavier ions would suffer break--up reactions on CMB
photons. The upshot is that no known UHE particle which interacts high in the
atmosphere (as required by observation) would be able to travel over distances
exceeding $\sim 50$ Mpc \cite{siglrev}. Intergalactic magnetic fields should
not be able to randomize the arrival directions of UHE particles over such
distances, i.e.~they should still (more or less) point back to their sources.
However, there are no known sources within this radius in the directions of
these UHE events.

This has lead to various exotic proposals \cite{siglrev,dreesrev}. In
particular, it has been suggested \cite{xold, siglrev} that UHE events
originate from the decay of very massive, long--lived particles $X$ with mass
$M_X \gsim 10^{12}$ GeV. These could be particles associated with a Grand
Unified theory, which are stabilized by being bound in topological defects
\cite{xold,neck}, or free particles with extremely small couplings to ordinary
matter \cite{freex}. In either case, they would have been produced in the very
early universe, just after the end of inflation \cite{xold,prodx}. The GZK
effect could then be circumvented, if most of these decays occur within one
GZK interaction length, for example in the halo of our own galaxy. The decay
of an $X$ particle triggers a parton cascade, followed by hadronization and
the decay of unstable particles \cite{xdec,bad}.

One potential problem of this scenario is that it tends to predict a higher
flux for UHE photons than protons, at least at the point of $X$ decay. For $E
\ll M_X$ this is due to the fact that fragmentation produces more pions than
protons, and neutral pions decay into photons; for $E \sim M_X/2$ the direct
emission of photons in the early stages of the parton cascade also plays a
role. This is problematic, since experiments indicate that most UHE events are
caused by protons (or heavier nuclei) rather than photons \cite{prot}.
However, the statistics at post--GZK energies is still quite poor, and our
knowledge of the details of interactions at $E > 10^{19}$ eV (corresponding to
$\sqrt{s} > 10^5$ GeV) still leaves much to be desired. Propagation effects
might also modify the proton to photon flux ratio. Finally, protons from
conventional (``bottom--up'') sources might contribute significantly (although
not dominantly) to the observed UHE events, thereby increasing the proton to
photon ratio \cite{ellis}. Given that other explanations also have their
problems \cite{dreesrev}, top--down models should still be considered
viable.

These models predict an UHE neutrino flux that is even higher than the photon
flux, since fragmentation produces more charged than neutral pions, and more
$W$ bosons than photons are emitted during the early stage of the parton
cascade. These models can therefore also be tested \cite{neutx} by looking for
neutrinos with $E > 10^5$ GeV, where the atmospheric neutrino background
becomes negligible. However, bottom--up models also predict a substantial UHE
neutrino flux, e.g. due to the GZK effect itself. Detailed analyses of the
spectra of neutrino events would therefore be required to distinguish between
top--down and bottom--up models.

There is, however, a potential ``smoking gun'' signature for top--down models.
The hierarchy between the scale $M_X \gsim 10^{12}$ GeV and the weak scale
will in general only be stable against radiative corrections in the presence
of weak--scale supersymmetry \cite{book}. If $R$ parity is conserved, the
lightest superparticle (LSP) is stable, and will also be produced copiously in
the decay of $X$ particles \cite{bk,bad}. In most supersymmetric models, the
best LSP candidate (in the visible sector) is the lightest neutralino $\tilde
\chi_1^0$ \cite{book}. Early estimates \cite{bk} indicated that the cross
section for neutralino interactions with ordinary matter is significantly
smaller than that of neutrinos. This lead to the suggestion
\cite{bdhh2,luis,mele} to search for UHE neutralinos by using the Earth as a
filter: for some range of energies the neutrino spectrum would get depleted in
the Earth, but neutralinos would still come through, and might be detectable
by future experiments.

Clearly a good understanding of the interactions of UHE neutralinos with
matter is mandatory in order to analyze the viability of this signal. This is
the topic of our paper. The propagation of neutralinos through the Earth was
treated using a simple Monte Carlo model in ref.\cite{bdhh2}, whereas
ref.\cite{luis} ignored all neutralinos that had any interactions in the Earth
(as opposed to in the detection medium). This is ``overkill'', since for
models with conserved $R$ parity, any neutralino interaction will eventually
result in another LSP, i.e.~another neutralino, albeit with reduced energy. In
this regard neutralinos are like $\tau$ neutrinos, which are also regenerated
after interacting with ordinary matter \cite{nutau,reno1}. In order to treat
this effect, one has to know the interaction cross section differential in the
scaling variable $y \equiv 1 - E_{\rm out} / E_{\rm in}$, where $E_{\rm in}$
and $E_{\rm out}$ are the energies of the incoming and outgoing neutralino,
respectively. The calculation of this cross section is described in Sec.~2,
including for the first time the case of higgsino--like neutralino. In Sec.~3
we describe neutralino propagation through matter by means of transport
equations, similar to the ones used to describe $\nu_\tau$ propagation
\cite{reno1}. We find that an iterative solution of this equation along the
lines of refs.\cite{iter,reno1,rr} is not applicable in our case, since it
leads to a significant violation of the conservation of the total neutralino
flux; this flux conservation is a direct consequence of the regeneration
mechanism described above. We therefore use a straightforward numerical
integration of the transport equations. Finally, Sec.~4 contains a brief
summary and some conclusions. The calculation of a new, usually subdominant
contribution to the neutralino nucleon scattering cross section is described
in an Appendix.

\section{Calculation of cross sections}

This Section deals with the calculation of the total and differential cross
section for the scattering of ultra--relativistic neutralinos $\lsp$ on
nuclei. This process dominates by far the total interaction of neutralinos
with matter, except for a very narrow range of energies around $E_{\tilde
  \chi_1^0} = (m^2_{\tilde e} - m^2_{\tilde \chi_1^0}) / (2 m_e)$, where
on--shell selectrons $\tilde e$ can be produced in $\tilde \chi_1^0 e$
scattering \cite{mele}. In the following two subsections we will discuss
$s-$channel and $t-$channel scattering, respectively; we will see that the
former (latter) process dominates for bino-- (higgsino--)like LSP.

\subsection{$s-$channel contribution}

The $s-$channel contribution to neutralino--quark scattering, $\lsp q_i
\rightarrow X$, is described by the Feynman diagram shown in Fig.~1.
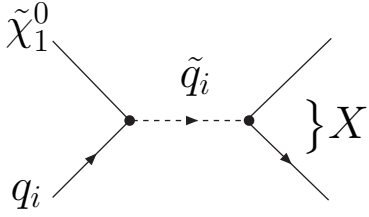
\begin{figure}
\begin{center}
\begin{picture}(169,127) (209,-151)
\Text(210,-73)[lb]{\Large{$\tilde{\chi}_1^0$}}
\Text(212,-133)[lb]{\Large{$q_i$}}
\Text(322,-112)[lb]{\Large{$\bigg\}X$}}
\Text(276,-90)[lb]{\Large{$\tilde{q_i}$}}
\SetWidth{0.5}
\DashArrowLine(257,-99)(302,-99){2}
\Line(301,-98)(332,-68)
\ArrowLine(300,-99)(331,-129)
\ArrowLine(227,-128)(257,-98)
\Line(227,-69)(257,-100)
\Vertex(256,-99){2.0}
\Vertex(301,-99){2.0}
\end{picture}
\caption{\label{fig1}$s-$channel Feynman diagram for $\lsp q_i$ scattering,
  where $\tilde{q_i}$ is a virtual squark and the symbol $X$ stands for all
  allowed final states. The arrows indicate the flow of baryon number. An
  analogous diagram exists for $\lsp \bar q_i$ scattering, with reversed
  arrows.}
\end{center}
\end{figure} 
The total partonic cross section can be written as follows:
\eqa
\hat\sigma(\hat s) = \pi \frac{1} {|\vec P_1^*|^2} \frac{\hat s}
 {(\hat s-m_{\tilde q_i}^2)^2 + m_{\tilde q_i}^2 \Gamma_{\tilde{q_i}}^2}
  \Gamma( \tilde q_i \rightarrow q_i + \lsp)
  \Gamma_{\tilde q_i}
  \hspace{0.1cm} \textrm{,} \label{ansatzskanal}
\eqe
where $\hat s$ is the partonic center--of--mass (c.m.) energy, $m_{\tilde
  q_i}$ is the squark mass, $\vec P_1^* = (\hat s - m^2_{\tilde \chi_1^0}) / (
2 \sqrt{\hat s})$ is the c.m. 3--momentum of the incoming particles,
$\Gamma_{\tilde q_i}$ is the total decay width of the squark and
$\Gamma(\tilde q_i \rightarrow q_i + \lsp)$ is the partial $\tilde q_i
  \rightarrow q_i + \lsp$ decay width. The same expression also holds for
$\lsp \bar q_i$ scattering.

The $s-$channel contribution to the total $\lsp-$nucleon scattering cross
section can be evaluated from Eq.(\ref{ansatzskanal}) by convoluting with the
appropriate (anti--)quark distribution function and summing over flavors. We
use the narrow width approximation,
\begin{equation} \label{narrow}
\frac{1} { \kll \hat s - m^2 \klr ^2 + m^2\Gamma^2} \stackrel{\Gamma
  \rightarrow 0} {\longrightarrow}  \frac{\pi} {m \Gamma} \delta\!(\hat s -
  m^2) \, .
\end{equation} 
The convolutions then collapse to simple products \cite{luis}:
\eqa  \label{es} 
\sigma^{\rm tot}_s = \frac {\pi} {4} \sum_{q} \kll | a_{q_L} |^2 + | a_{q_R}
     |^2 \klr \frac {1} {m^2_{\tilde q} } x q(x,Q^2) \, ,
\eqe
with
\begin{equation} \label{ex}
x = \frac {m^2_{\tilde q} - m^2_{\tilde \chi_1^0}} {2m_N E_{\rm in}}\, ,
\end{equation}
where $E_{\rm in}$ is the energy of the incident neutralino in the rest frame
of the nucleon, and $m_N = (m_p + m_n)/2$ is the nucleon mass. In our
numerical evaluation we use $Q^2 = m^2_{\tilde q}$ as momentum scale in the
(anti--)quark distribution functions. For simplicity we assume equal masses
$m_{\tilde q}$ for the $L$ and $R$ squarks of a given flavor. In this case
left-- and right--handed couplings contribute symmetrically, as shown in
Eq.(\ref{es}).\footnote{In general $a_{q_L}$ only contribute to $\tilde q_L$
exchange, while $a_{q_R}$ contribute to $\tilde q_R$ exchange.} Our simplified
treatment is sufficient as long as squark masses remain free parameters; note
also that most SUSY models predict small mass splittings between squarks, at
least for the first two generations \cite{book}. Finally, the couplings
appearing in Eq.(\ref{es}) are given by:
\eqa
a_{q_L} &=& \sqrt{2} g_2 \left( T_{3,q} N_{12} + \frac {\tan\theta_W} {6}
  N_{11} \right) \ \ \ {\rm for} \ q = u, d, s, c\, ; \nonumber \\
a_{q_R} &=& \sqrt{2} g_2 \tan\theta_W Q_q  N_{11}  \ \ \ {\rm for} \ q = u, d,
s, c \, ;       \nonumber   \\
a_{b_{L,R}} &=& a_{d_{L,R}} + \frac{g_2 m_b}{\sqrt{2} m_W \cos\beta} N_{13}
\hspace{0.1cm} \textrm{.}
\label{defcoup}
\eqe
Here, $N_{ij}$ are the entries of the neutralino mixing matrix in the notation
of ref.\cite{gh86}, $g_2$ is the $SU(2)$ coupling constant, $\theta_W$ is the
weak mixing angle, $T_{3,u} = - T_{3,d} = 1/2$ is the weak isospin, $Q_q$ is
the electric charge of quark $q$ in units of the proton charge, $m_W$ is the
mass of the $W^{\pm}$ boson, $\tan\beta$ is the ratio of the Higgs vacuum
expectation values, and $m_b$ is the mass of the bottom quark. We ignore the
masses of quarks of the first and second generation in these couplings,
i.e.~we use identical couplings for up and charm quarks, as well as for down
and strange quarks. We do include contributions $\propto m_b$ to the couplings
of the bottom quark \cite{gh86}.

Note that we do not include the contribution from top (s)quarks in
Eq.(\ref{es}). Since the top quark may not be much lighter than their
superpartners, it is more appropriate to treat $\tilde t$ production through
the $2 \rightarrow 2$ scattering reactions
\begin{equation} \label{etop}
\lsp g \rightarrow \tilde t \bar t, \ \bar{\tilde t} t\, .
\end{equation}
We evaluated the corresponding cross section, but found it to be subdominant
in all scenarios we considered; this is not very surprising, since it is of
higher order in the strong coupling than the cross section (\ref{es}). We
therefore delegate the discussion of the reaction (\ref{etop}) to the
Appendix.\footnote{One should not add the cross section from $\lsp g
  \rightarrow \bar q \tilde q$ to the cross section (\ref{es}), as is done in
  ref.\cite{bk}. The reason is that the result (\ref{es}) already includes
  contributions from $g \lsp$ scattering in the leading logarithmic
  approximation, via $g \rightarrow q \bar q$ splitting, which contributes to
  the $Q^2$ dependence of the quark distribution functions. Simply adding both
  contributions therefore leads to double counting. Instead, $g \lsp$
  scattering should be treated as part of higher--order corrections to
  inclusive $\tilde q$ production described by Eq.(\ref{es}), {\em provided}
  the quark $q$ can be assigned a partonic distribution function in the
  nucleon.}

In order to define potentially realistic supersymmetric scenarios, we work in
the framework of minimal supergravity (mSUGRA) \cite{book}, which is the
simplest supersymmetric model generically yielding a stable neutralino as LSP
(assuming that the gravitino is heavier than the $\lsp$). Here one postulates
universal sfermion ($m_0$) and gaugino masses ($m_{1/2}$) at the scale of
Grand Unification.  We use the public code Softsusy \cite{All02} to calculate
spectra for three representative scenarios yielding a bino--like state as LSP,
as defined in Table~\ref{tab1}. For simplicity we fix the trilinear
interaction parameter $A_0 = 0$ and $\tan\beta = 2$. We choose rather small or
moderate values of the scalar and gaugino mass parameters; larger masses yield
smaller cross sections, and hence smaller effects from propagation through
Earth.\footnote{Within mSUGRA, the combination of small $\tan\beta$ and small
  soft breaking masses is excluded by Higgs searches at LEP. Note, however,
  that $\tan\beta$, or indeed the structure of the Higgs sector as a whole,
  plays little role in our analysis. Scenarios with rather light squarks and
  gauginos are still allowed in other supersymmetric models. The squark masses
  in scenarios D2 and D3 are well above the lower bound derived in \cite{ddk3}
  in the framework of mSUGRA.}

\begin{table}[h!]
\begin{center}
\begin{tabular}{|c|c|c|c|c|} 
\hline
\multicolumn{5}{|c|}{\bf mSUGRA scenarios} \\
\hline
Scenario & $m_0$ &  $m_{1/2}$ &  $m_{\lsp}$ & $m_{\tilde d_L}$ \\
\hline
\hline
D1 & 80 & 150 & 63 & 365 \\ 
\hline
D2 & 150 & 250 & 104 & 582 \\ 
\hline
D3 & 250 & 450 & 189 & 992 \\ 
\hline
\end{tabular}
{\caption{\label{tab1}mSUGRA scenarios for $\tan \beta=2$, higgsino mass
    parameter $\mu < 0$ and $A_0=0$. $m_0$ and $m_{1/2}$ are the universal
    scalar and gaugino mass parameters, respectively. $m_{\tilde{d}_L}$ is the
    mass of the $SU(2)$ doublet down squark, but all other squark masses have
    quite similar values. All masses are in GeV.}}
\end{center} 
\end{table} 

Results for the corresponding total cross sections for scattering on nucleons,
with and without the contribution from bottom (s)quarks, are presented in
Fig.~\ref{fig2}. Here and in the following figures we use the CTEQ6
parameterization \cite{KLOT03} of the parton distribution functions, averaged
over proton and neutron targets; in the absence of large mass splitting
between $\tilde u$ nd $\tilde d$ squarks, the cross sections for scattering on
protons and neutrons are very similar. We see that the contribution of bottom
(s)quarks only reaches 10\% even for large squark masses and large $\lsp$
energy. In this case one is probing the $b-$quark distribution function at
small Bjorken$-x$, see Eq.(\ref{ex}), and rather high $Q^2$, where it is
comparable to the distribution functions of light quarks. However, the small
hypercharge of bottom quarks suppresses their couplings to bino--like
neutralinos, relative to those of up and charm quarks; see
Eq.(\ref{defcoup}).

\begin{figure}[h!]
\begin{center}
\includegraphics[width=12cm]{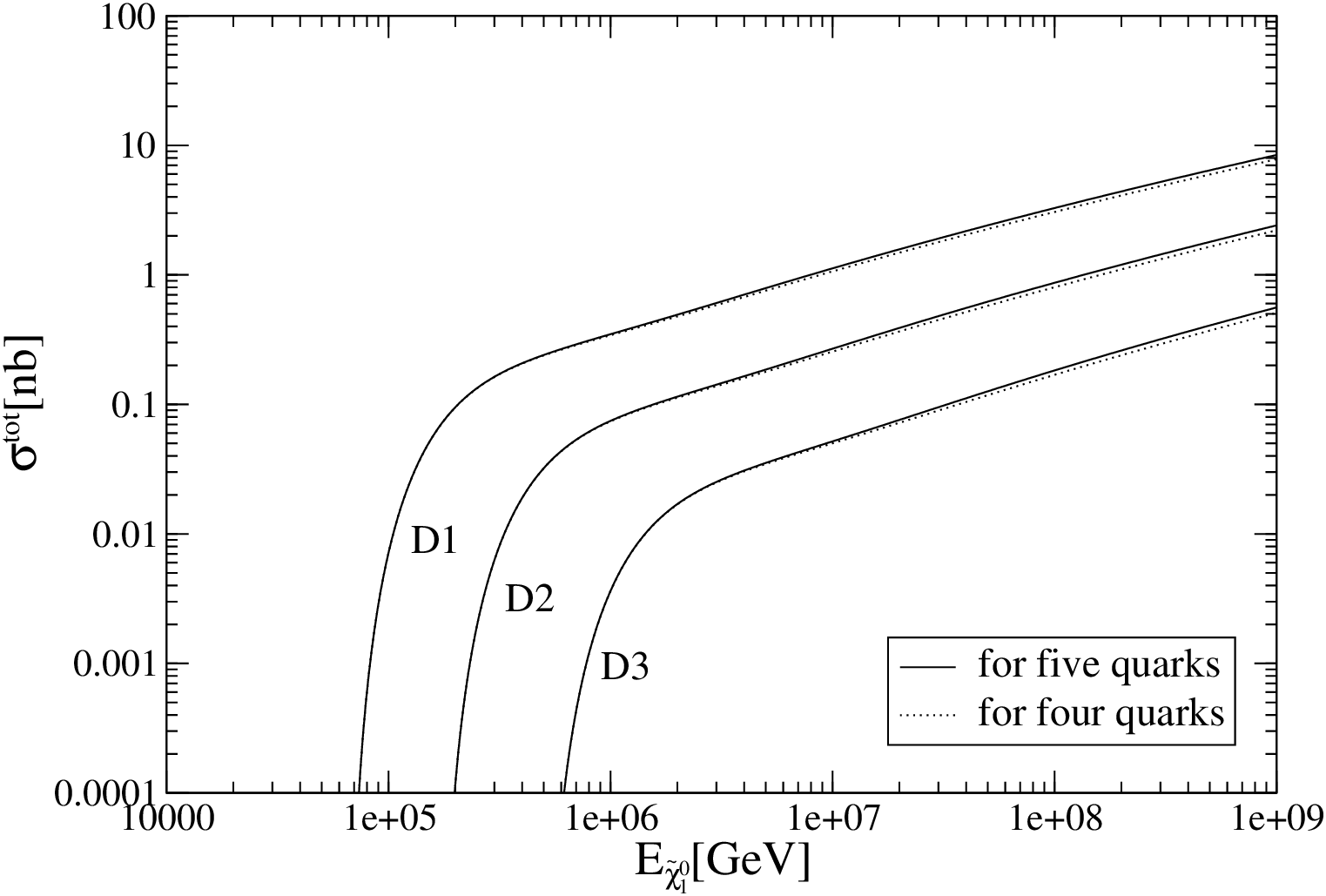}
\caption{\label{fig2}The total cross section of the $s-$channel contributions
  to $\lsp-$nucleon scattering for four (dotted curves) and five (solid
  curves) active quark flavors. The supersymmetric scenarios D1, D2 and D3 are
  defined in Table~\ref{tab1}.}
\end{center}
\end{figure}

Over most of the parameter space of mSUGRA the lightest neutralino is bino
dominated \cite{book,ddk3}; this is also true for the three scenarios of
Table~\ref{tab1}. Since the bino coupling to a sfermion is proportional to the
hypercharge of that sfermion, a bino--like $\lsp$ couples predominantly to
$SU(2)$ singlet, ``right--handed'' squarks, whose hypercharges are two (for
$\tilde d_R$) or four (for $\tilde u_R$) times larger than that of the $SU(2)$
doublet squarks; see Eqs.(\ref{defcoup}). The total $s-$channel cross section
is therefore dominated by the production of $SU(2)$ singlet squarks. Since
these squarks do not couple to $SU(2)$ gauginos, they will decay directly into
$q + \lsp$, if the gluino is heavier than these squarks \cite{bbkt}; in mSUGRA
this corresponds to $m_0 \lsim m_{1/2}$. In this case we can approximate the
total $s-$channel contribution as production of on--shell squarks which decay
back into $q + \lsp$ final states.

This greatly simplifies the calculation of the cross section differential in
the scaling variable $y \equiv 1 - E_{\rm out} / E_{\rm in}$, where $E_{\rm
  in}$ and $E_{\rm out}$ are the incoming and outgoing $\lsp$ energy in the
nucleon rest frame. The crucial observation is that squark decays are
isotropic in the squark rest frame, which implies
\eqa
\frac {d\!\sigma_s} {d\!\cth^{\ast} } = \frac {\sigma_s^{\rm tot}} {2} \, ,
\eqe
where $\theta^*$ is the angle between the ingoing and outgoing $\lsp$ in this
frame. In order to get the final expression for the $y$ distribution, we have
to boost from the c.m. system into the rest frame of the nucleon. This yields
a flat distribution,
\eqa \label{diffxs}
\frac{d\!\sigma_s}{d\!y} = \frac{\sigma_s^{\rm tot}} {y_{\rm max}}
\, , \eqe
where
\eqa \label{ymax}
y_{\rm max} &=& 1 - \frac {m^2_{\tilde \chi_1^0}} {m^2_{\tilde q}} \, ;
\nonumber \\
y_{\rm min} &=& 0\, .
\eqe
In the first Eq.(\ref{ymax}) we have used $\hat s = m^2_{\tilde q}$ for
on--shell squark production. Forward scattering in the squark rest frame leads
to $E_{\rm out} = E_{\rm in}$, independent of the details of the kinematics;
the lower limit for $y$ is therefore always zero. On the other hand, $y=1$,
which requires $E_{\rm out} = 0$, is possible only for $\mlsp \rightarrow 0$.
Since in mSUGRA $m^2_{\tilde \chi_1^0} \ll m^2_{\tilde q}$, $y_{\rm max}$ is
indeed quite close to unity. An LSP will then lose on average about half its
energy if it undergoes $s-$channel scattering on a nucleon.

\subsection{$t-$channel contribution}

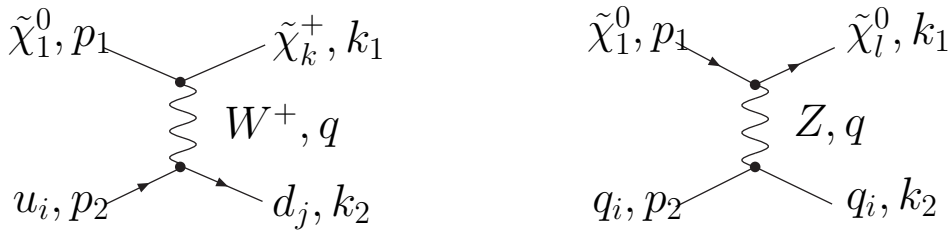
\begin{figure} 
\hspace{1.5cm}
\begin{minipage}[c]{5.0cm}
\begin{center}
\begin{picture}(152,107) (285,-177)
\SetWidth{0.5}
\Photon(345,-148)(345,-118){5}{3}
\Line(344,-117)(376,-103)
\Line(315,-104)(345,-117)
\ArrowLine(316,-163)(345,-149)
\ArrowLine(345,-149)(375,-162)
\Text(280,-108)[lb]{\Large{$\lsp,p_1$}}
\Text(282,-170)[lb]{\Large{$u_i,p_2$}}
\Text(380,-111)[lb]{\Large{$\cpk,k_1$}}
\Text(380,-173)[lb]{\Large{$d_j,k_2$}}
\Text(362,-142)[lb]{\Large{$W^+,q$}}
\Vertex(344,-117){2.0}
\Vertex(344,-149){2.0}
\end{picture}
\end{center}

\end{minipage} 
\hspace{1.0cm}
\hspace{1.2cm}
\raisebox{0.23cm}{ 
\begin{minipage}[c]{5.0cm}

\begin{center}
\begin{picture}(121,119) (360,-179)
\SetWidth{0.5}
\Photon(405,-121)(405,-152){5}{3}
\Line(375,-167)(405,-152)
\Line(405,-152)(434,-166)
\ArrowLine(405,-121)(435,-107)
\ArrowLine(375,-105)(405,-121)
\Vertex(405,-121){2.0}
\Vertex(405,-152){2.0}
\Text(343,-110)[lb]{\Large{$\lsp,p_1$}}
\Text(345,-173)[lb]{\Large{$q_i,p_2$}}
\Text(441,-172)[lb]{\Large{$q_i,k_2$}}
\Text(441,-111)[lb]{\Large{$\nel,k_1$}}
\Text(421,-144)[lb]{\Large{$Z,q$}}
\end{picture}
\end{center}

\end{minipage}
}
\caption{\label{fig3}Feynman diagrams for $t-$channel neutralino--nucleon
  scattering. The arrows indicate the flow of baryon number, and $p_{1,2}, \,
  k_{1,2}$ and $q$ refer to the four--momenta of the initial, final and
  exchanged particles, respectively.}
\end{figure}

We now turn to the calculation of the $t-$channel contribution to the LSP
nucleon scattering cross section. As shown in Fig.~\ref{fig3}, there are both
$W$ (charged current) and $Z$ exchange (neutral current) diagrams; the former
produce an outgoing chargino, while the latter have one of the four
neutralinos in the final state. There are additional diagrams with antiquarks
in both initial and final state, as well as charged current diagrams producing
a negative chargino from scattering off a $d$ or $\bar u$ quark.
Table~\ref{tab2} shows all possible initial and final states, where we again
include five active flavors of quarks in the nucleon.

\begin{table}[h] 
\begin{center}
\begin{tabular}{|ccc|c||ccc|c|} 
\hline
\multicolumn{4}{|c||}{{\bf $W^{\pm}$ exchange }} &
\multicolumn{4}{c|}{{\bf $Z$ exchange}} \\
\hline
\hline
\multicolumn{3}{|c|}{{Process }} & \# &
\multicolumn{3}{|c|}{{Process }} & \#  \\
\hline
\hline
$\lsp$ &   & $\cpk$ &  & $\lsp$ &   & $\nel$ &  \\ 
$u$    &\raisebox{1.5ex}[-1.5ex]{$\rightarrow$}& $d$ &
\raisebox{1.5ex}[-1.5ex]{$2$} &  
$u$    &\raisebox{1.5ex}[-1.5ex]{$\rightarrow$}& $u$ &
\raisebox{1.5ex}[-1.5ex]{$4$} \\  
\hline
$\lsp$ &   & $\cpk$ &  & $\lsp$ &   & $\nel$ &  \\ 
$c$    &\raisebox{1.5ex}[-1.5ex]{$\rightarrow$}& $s$ &
\raisebox{1.5ex}[-1.5ex]{$2$} &  
$c$    &\raisebox{1.5ex}[-1.5ex]{$\rightarrow$}& $c$ &
\raisebox{1.5ex}[-1.5ex]{$4$} \\ 
\hline
$\lsp$ &   & $\cmk$ &  & $\lsp$ &   & $\nel$ &  \\ 
$\bar{u}$    &\raisebox{1.5ex}[-1.5ex]{$\rightarrow$}& $\bar{d}$ &
\raisebox{1.5ex}[-1.5ex]{$2$} &  
$\bar{u}$    &\raisebox{1.5ex}[-1.5ex]{$\rightarrow$}& $\bar{u}$ &
\raisebox{1.5ex}[-1.5ex]{$4$} \\  
\hline
$\lsp$ &   & $\cmk$ &  & $\lsp$ &   & $\nel$ &  \\ 
$\bar{c}$    &\raisebox{1.5ex}[-1.5ex]{$\rightarrow$}& $\bar{s}$ &
\raisebox{1.5ex}[-1.5ex]{$2$} &  
$\bar{c}$    &\raisebox{1.5ex}[-1.5ex]{$\rightarrow$}& $\bar{c}$ &
\raisebox{1.5ex}[-1.5ex]{$4$} \\  
\hline
$\lsp$ &   & $\cpk$ &  & $\lsp$ &   & $\nel$ &  \\ 
$\bar{d}$    &\raisebox{1.5ex}[-1.5ex]{$\rightarrow$}& $\bar{u}$ &
\raisebox{1.5ex}[-1.5ex]{$2$} &  
$\bar{d}$    &\raisebox{1.5ex}[-1.5ex]{$\rightarrow$}& $\bar{d}$ &
\raisebox{1.5ex}[-1.5ex]{$4$} \\  
\hline
$\lsp$ &   & $\cpk$ &  & $\lsp$ &   & $\nel$ &  \\ 
$\bar{s}$    &\raisebox{1.5ex}[-1.5ex]{$\rightarrow$}& $\bar{c}$ &
\raisebox{1.5ex}[-1.5ex]{$2$} &  
$\bar{s}$    &\raisebox{1.5ex}[-1.5ex]{$\rightarrow$}& $\bar{s}$ &
\raisebox{1.5ex}[-1.5ex]{$4$} \\  
\hline
$\lsp$ &   & $\cpk$ &  & $\lsp$ &   & $\nel$ &  \\ 
$\bar{b}$    &\raisebox{1.5ex}[-1.5ex]{$\rightarrow$}& $\bar{t}$ &
\raisebox{1.5ex}[-1.5ex]{$2$} &  
$\bar{b}$    &\raisebox{1.5ex}[-1.5ex]{$\rightarrow$}& $\bar{b}$ &
\raisebox{1.5ex}[-1.5ex]{$4$} \\   
\hline
$\lsp$ &   & $\cmk$ &  & $\lsp$ &   & $\nel$ &  \\ 
$d$    &\raisebox{1.5ex}[-1.5ex]{$\rightarrow$}& $u$ &
\raisebox{1.5ex}[-1.5ex]{$2$} &  
$d$    &\raisebox{1.5ex}[-1.5ex]{$\rightarrow$}& $d$ &
\raisebox{1.5ex}[-1.5ex]{$4$} \\  
\hline
$\lsp$ &   & $\cmk$ &  & $\lsp$ &   & $\nel$ &  \\ 
$s$    &\raisebox{1.5ex}[-1.5ex]{$\rightarrow$}& $c$ &
\raisebox{1.5ex}[-1.5ex]{$2$} &  
$s$    &\raisebox{1.5ex}[-1.5ex]{$\rightarrow$}& $s$ &
\raisebox{1.5ex}[-1.5ex]{$4$} \\  
\hline
$\lsp$ &   & $\cmk$ &  & $\lsp$ &   & $\nel$ &  \\ 
$b$    &\raisebox{1.5ex}[-1.5ex]{$\rightarrow$}& $t$ &
\raisebox{1.5ex}[-1.5ex]{$2$} &  
$b$    &\raisebox{1.5ex}[-1.5ex]{$\rightarrow$}& $b$ &
\raisebox{1.5ex}[-1.5ex]{$4$} \\  
\hline
\hline
 & & $\sum$ & $20$  &  &  &$\sum$  & $40$   \\
\hline
\end{tabular}
\caption{\label{tab2}List of processes contributing to $\lsp-$nucleon
  scattering through the exchange of an electroweak gauge boson in the
  $t-$channel, together with the number of different final states.  The
  chargino index $k$ runs from 1 to 2 and the neutralino index $l$ from 1 to
  4. }
\end{center}
\end{table}

The partonic total cross section can be obtained by integrating over the
scattering angle $\theta^*$ in the center--of--mass frame; convolution with
the relevant quark distribution functions then yields the $t-$channel
contribution to the $\lsp-$nucleon scattering cross section:
\eqa
\sigma_t^{\rm tot}(s) &=& \sum_q \int_{x_{\rm min}}^1 dx
\int_{-1}^{\cth^*_{\rm max}}
d\!\cth^*  \frac {1} {32\pi} q(x,Q^2) \frac{| \mathcal{M} |^2 |\vec{P}^*_2 \kll
  \hat{s} \klr |} {\hat{s}| \vec{P}_1^* \kll \hat{s} \klr| } \, ,
\label{ttot} 
\eqe
where $\hat{s} = 2 x E_{\rm in} m_N + m^2_{\tilde \chi_1^0}$, and
$\vec{P}_1^*$ and $\vec{P}^*_2$ are the three--momenta of the incoming and
outgoing particles in the c.m. system, respectively. As usually done in
deep--inelastic scattering, we identify the scale $Q^2$ in the quark
distribution functions with the absolute value of the four--momentum $\hat t$
exchanged between the participating partons. This causes a minor difficulty:
forward scattering ($\cth^* = +1$) leads to $Q^2 = \hat t = 0$, where the
parton distribution functions are not defined. Demanding $Q^2 > Q^2_{\rm min}$
therefore leads to a restriction on the phase space integration:
\begin{equation} \label{cmax}
\cth^*_{\rm max} = 1 - \frac {2 \hat s Q^2_{\rm min}} { \left( \hat s -
    m^2_{\tilde \chi_1^0} \right) \left( \hat s - m^2_{\tilde \chi_{\rm out}}
    \right) }\, ,
\end{equation}
where $\tilde \chi_{\rm out}$ stands for the chargino or neutralino in the
final state. This in turn affects the lower bound on the momentum fraction $x$:
\eqa \label{xmin}
x_{\rm min} &=& \frac{ \frac{1}{2}
 \kll Q_{\rm min}^2 + m_{\tilde{\chi}_{\rm out}}^2 -m_{\lsp}^2 \klr +
 \sqrt{\frac{1}{4} \kll Q^2_{\rm min} + m_{\tilde{\chi}_{\rm out}}^2 +
   m_{\lsp}^2 \klr^2 - m_{\tilde{\chi}_{\rm out}}^2 m_{\lsp}^2} }
 {2E_{\rm in} m_N} \, ,
\eqe
where $E_{\rm in}$ is again the energy of the incoming neutralino in the
nucleon rest frame. The total cross section depends only very weakly on the
choice of $Q_{\min}^2$. The $W$ and $Z$ propagators become independent of
$Q^2 = |\hat t|$ once $Q^2 \ll m_W^2$. Moreover, as we will see shortly, the
total cross section is dominated by the production of $\tilde \chi_{\rm out}
\neq \lsp$, where $Q_{\rm min}^2 \ll m^2_{\tilde \chi_{\rm out}}$ only has a
very small effect on $x_{\rm min}$. In our numerical calculations we fixed
$Q^2_{\rm min} = 1$ GeV$^2$.

The squared matrix elements $|\mathcal{ M}|^2$ for the charged current
reactions are given by 
\eqa
\frac{1}{4}\sum_{\rm spins} |\mathcal{ M}|^2
 &=& \frac{g_2^4} {\kll \hat t - M_W^2 \klr^2} \Bigl[ 2|\cei|^2 \pepz\:\kekz + 
 2|\czw|^2\pekz\:\pzke \nonumber \\
 &&\hspace*{2cm} - \kll \ceis \czw + \cei \czws \klr \pzkz \mcpk \mnel \Bigr] 
\, , \label{tm1}
\eqe
for the first four cases of Table~\ref{tab2}, and 
\eqa
\frac{1}{4} \sum_{\rm spins } |\mathcal{ M}|^2
 &=& \frac{g_2^4} {\kll \hat t - M_W^2 \klr^2} \bigl[
  2|\cei|^2\pekz\:\pzke+  2|\czw|^2\pepz\:\kekz  \nonumber \\
 &&\hspace*{2cm} 
-\kll \ceis \czw + \cei \czws \klr \pzkz \mcpk \mnel \bigr] \, , \label{tm2}
\eqe
for the last six cases of Table~\ref{tab2}; note that the results (\ref{tm1})
and (\ref{tm2}) differ by the exchange of the left-- and right--handed $W^\pm
\lsp \tilde \chi_k^\mp$ couplings. These couplings are given by \cite{gh86}
\eqa \label{ccoup}
\cei &=& N_{12} \mathcal{V}_{k1}^{\ast} - \frac {1} {\sqrt{2}} N_{14}
\mathcal{V}_{k2}^{\ast} \, , \nonumber \\
\czw &=& N_{12}^{\ast} \mathcal{U}_{k1} + \frac {1} {\sqrt{2}}N_{13}^{\ast}
    \mathcal{U}_{k2}  \, .
\eqe
Here, the index $k$ runs from 1 to 2, $\mathcal{U}$ and $\mathcal{V}$ denote
the chargino mixing matrices, $\mcpk$ is the mass of the outgoing chargino,
and $\hat t = -Q^2 = (k_1 - p_1)^2$; the four--momenta have been defined in
Fig.~\ref{fig3}. In Eqs.(\ref{tm1}) and (\ref{tm2}) we have ignored quark
flavor mixing, i.e.~we replaced the quark mixing matrix by the unit matrix.

Due to the Majorana nature of the neutralinos, neutral current reactions on a
quark and antiquark are described by the same matrix element:
\eqa
\frac {1} {4} \sum_{\rm spins} |\mathcal{ M}|^2
 &=& \frac {2} {c_W^4} \frac {g_2^4 \kll |\kei|^2 + |\kzw|^2 \klr} {\kll t -
   M_W^2 \klr^2} \Bigl[ \kll |\nei|^2 +|\nzw|^2 \klr \kll \pepz\:\kekz
 + \pekz \:\pzke \klr \nonumber \\
 && \hspace*{3.4cm} - \kll \nei \nzws + \nzw \neis
 \klr \pzkz \mlsp \mnel \Bigr] \, ,
 \label{tm3} 
\eqe
with
\eqa \label{ncoup}
c_W&=&\cos\theta_W \, , \nonumber \\
\nei &=& \frac {1} {2} \kll - N_{l3} N_{13}^{\ast} + N_{l4} N_{14}^{\ast} \klr
\, , \nonumber \\ 
\nzw &=& - \kll \nei \klr^{\ast} \, , \nonumber \\
\kei &=&  T_{3,q} -  Q_q \sin^2 \theta_W \, , \nonumber \\ 
\kzw &=& Q_q \sin^2 \theta_W \, .  
\eqe
The index $l$ in Eqs.(\ref{tm3}) and (\ref{ncoup}) runs from $1$ to $4$, and
$Q_q$ and $T_{3,q}$  have been introduced in Eqs.(\ref{defcoup}).

\vspace*{5mm}
\begin{table}[h!] 
\begin{center}
\begin{tabular}{|c|c|c|c|c|c|c||c|}
\hline
Scenario & $m_{\lsp}$ & $m_{\tilde{\chi}_2^0}$ & $m_{\tilde{\chi}_3^0}$ &
$m_{\tilde{\chi}_4^0}$ & $m_{ \tilde{\chi}_1^{+}}$ & $m_{ \tilde{\chi}_2^{+}}$
& $GF$ [$\%$] \\
 \hline
D2 & 104 & 206 & 468 & 477 & 206 & 476 & 99.4 \\
 \hline
D3 & 189  & 367 & 800 & 805 & 367 & 805 & 99.8 \\
\hline
\hline
H1 & 125 & 137  & 742 & 801 & 130 & 742 & 1.2 \\
 \hline
H2 & 300 & 310 & 940 & 970 & 303 & 970 & 1.1 \\
 \hline
\end{tabular}
\caption{\label{tab3}Neutralino and chargino masses (in GeV) as well as the
  gaugino fraction $GF = |N_{11}|^2 + |N_{12}|^2$ for the four scenarios
  discussed in the text. Scenarios D2 and D3, where $\lsp$ is bino--like, have
  already been introduced in Table~\ref{tab1}, whereas scenarios H1 and H2
  describe higgsino--like $\lsp$; note that the higgsino fraction is $100\% -
  GF$.}
\end{center}
\end{table}

\begin{figure}[h!] 
\begin{center}
\includegraphics[width=12cm]{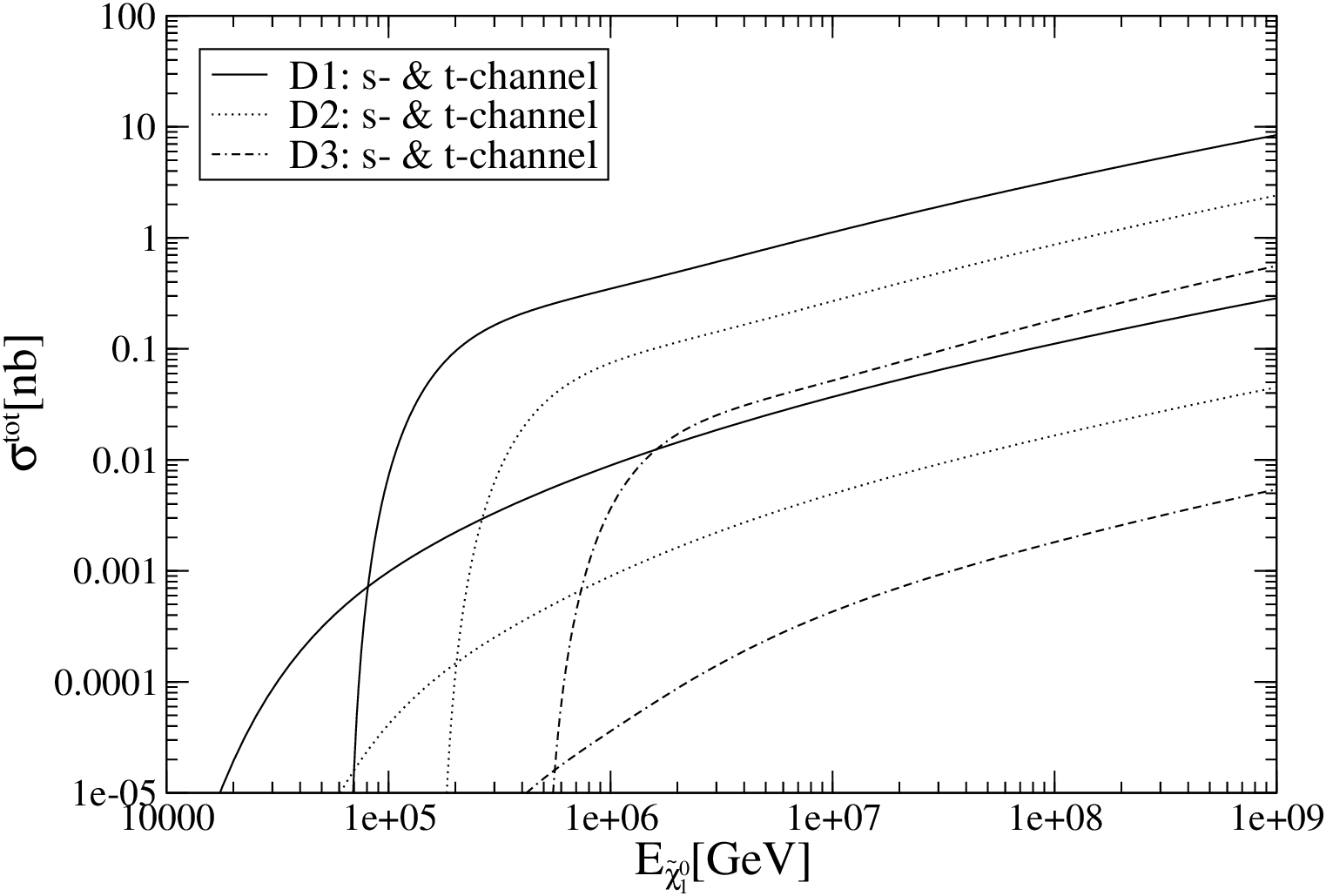}
\caption{\label{fig4}The total $\lsp-$nucleon scattering cross section for the
  three scenarios of Table~\ref{tab1}, where $\lsp$ is bino--dominated. At
  large energies the upper (lower) curves show the contribution from $s-$
  ($t-$)channel diagrams. Interference between these diagrams is negligible.}
\end{center}
\end{figure}

As mentioned earlier, all neutralino and chargino states can be produced in
$t-$channel $\lsp-$nucleon scattering. Table~\ref{tab3} lists the masses of
these particles for four scenarios. The first two have already been introduced
in Table~\ref{tab1} and describe bino--dominated $\lsp$ states, whereas in
scenarios H1 and H2 $\lsp$ is dominated by its higgsino components.
Eqs.(\ref{ccoup}) and (\ref{ncoup}) show that bino--like states, which have
$|N_{12}|, \, |N_{13}|, \, |N_{14}| \ll 1$, have suppressed couplings to gauge
bosons. We therefore expect the $t-$channel contributions to be subdominant
for the scenarios with bino--like $\lsp$. This is borne out by
Fig.~\ref{fig4}, which shows the $s-$ and $t-$channel contributions to the
total $\lsp-$nucleon scattering cross section for the three scenarios of
Table~\ref{tab1}. The $t-$channel contribution dominates only at low energies,
below the threshold for squark production; however, there the cross section is
in any case very small, and any possible signal from UHE LSPs will be masked
by the much larger neutrino signal. At higher energies the $s-$channel
contribution is at least a factor of 30 larger than those from all $t-$channel
diagrams. Note that the latter also decrease quickly with increasing sparticle
mass scale, i.e.~when going from scenario D1 over D2 to D3. The reason is that
increasing the gaugino mass parameter $m_{1/2}$ in mSUGRA not only increases
the bino mass, via the condition of electroweak symmetry breaking it also
increases the absolute value of the higgsino mass. Both effects tend to reduce
gaugino--higgsino mixing, and hence the couplings of $\lsp$ to gauge bosons.
These results show that one can indeed ignore all $t-$channel contributions
for bino--like LSPs, as done in refs.\cite{bdhh2,luis}.

\vspace*{10mm}

\begin{figure}[htb] 
\begin{center}
\includegraphics[width=12cm]{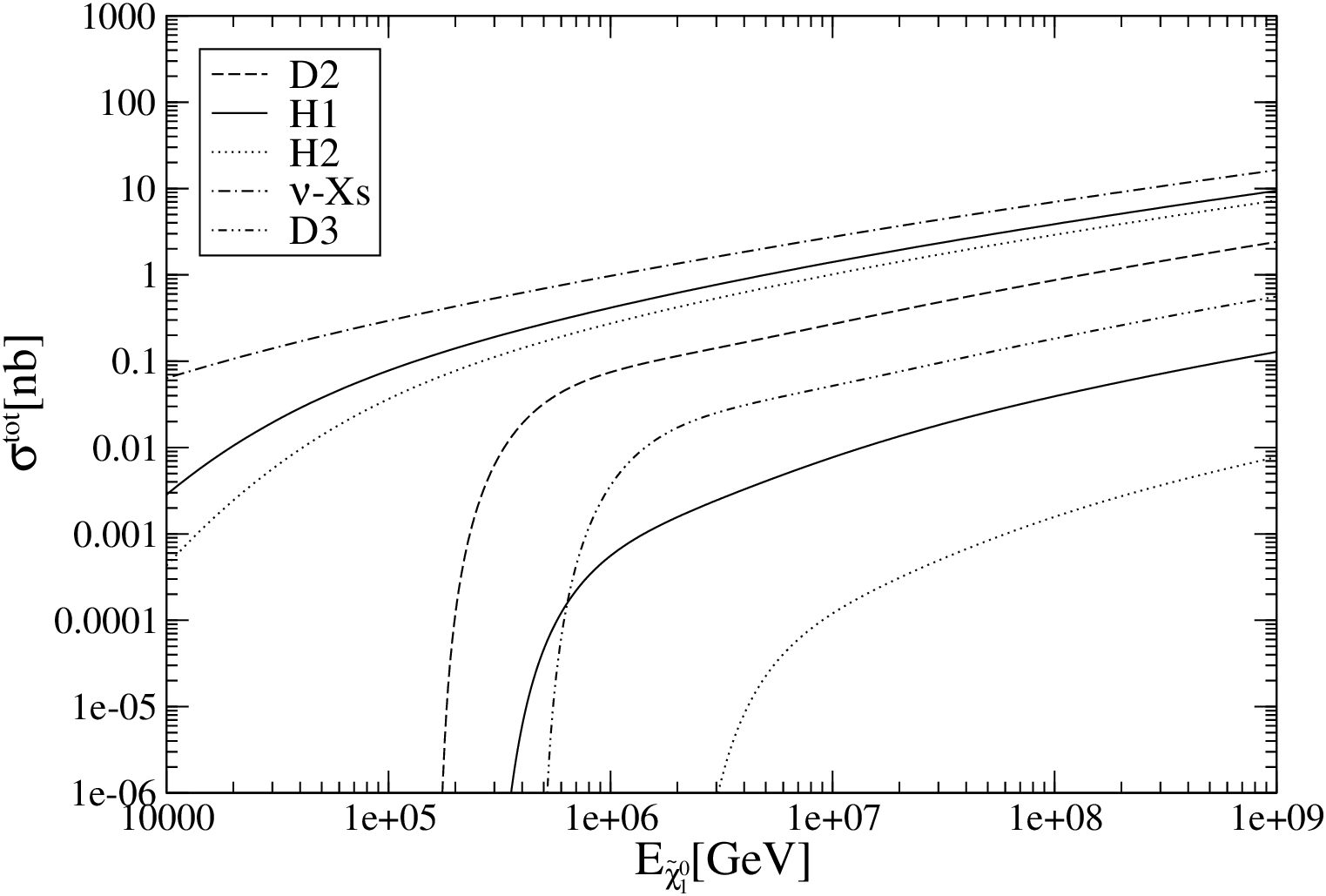}
\caption{\label{fig5}Comparison of total cross sections for neutrinos
  (dashed-dotted curve) and higgsino-- and bino--like $\lsp$. For scenarios H1
  and H2 (see Table~\ref{tab3}) with higgsino--like LSP, $t-$ and $s-$channel
  contributions are shown by the upper and lower curves, respectively; for
  scenarios D2 and D3, only the dominant $s-$channel contributions are shown.}
\end{center}
\end{figure}

In contrast, Eqs.(\ref{defcoup}) show that a higgsino--like $\lsp$, with
$|N_{11}|, \, |N_{12}| \ll 1$, has very small couplings to light quarks, which
suppresses the $s-$channel contribution, whereas $|N_{13}|, \, |N_{14}|$ are
sizable and lead to unsuppressed couplings to gauge bosons. We therefore
expect the $t-$channel contributions to dominate the total $\lsp-$nucleon
scattering cross section for higgsino--like $\lsp$. Fig.~\ref{fig5} shows that
this is indeed the case. The solid and dotted curves in this figure show the
total $t-$ and $s-$channel contributions in scenarios H1 and H2 of
Table~\ref{tab3}, where the $t-$channel contributions now correspond to the
{\em upper} curves. Here we used squark masses of around 1.9 TeV for scenario
H2, and about 1 TeV for H1; note that large scalar masses, $m_{\tilde q} \gg
|\mu|$ are required in mSUGRA if the LSP is to be higgsino--like \cite{book}.
For comparison we also show the total cross section for neutrino--nucleon
scattering \cite{GQRS98} (dot--dashed curve). Since neutrinos are also $SU(2)$
doublets, their cross section is similar to that of higgsino--like
neutralinos. However, some differences remain even at high energies. The
reason is that the quark distribution functions appearing in the expression
(\ref{ttot}) for the total $t-$channel contribution to the cross section peak
strongly at small $x$. As a result, the partonic cms energy $\sqrt{\hat s}$ is
often not much larger than the masses of the relevant charginos and
neutralinos, leading to some suppression of the LSP cross section relative to
the neutrino cross section. We will come back to this point shortly.

\vspace*{10mm}

The dashed and dash--doubledotted curves in Fig.~\ref{fig5} show again the
results for scenarios D2 and D3, where we have omitted the subdominant
$t-$channel contributions for clarity. We see that these scenarios lead to
somewhat smaller total cross sections than the scenarios with higgsino--like
LSP. The $s-$channel partonic subprocesses occur at first order in the weak
coupling, whereas the $t-$channel diagrams are of second order. However, this
relative enhancement of the $s-$channel contributions is over--compensated by
the fact that the relevant scale in the partonic $s-$channel cross section is
the squark mass, while for the $t-$channel cross sections it is given by the
mass of the exchanged gauge boson. On the other hand, comparison with
Fig.~\ref{fig4} shows that for scenario D1 with relatively light squarks, the
($s-$channel dominated) cross section is about the same as the ($t-$channel dominated)
cross section for higgsino--like LSP.

\begin{figure}[h!] 
\begin{center}
\includegraphics[width=12cm]{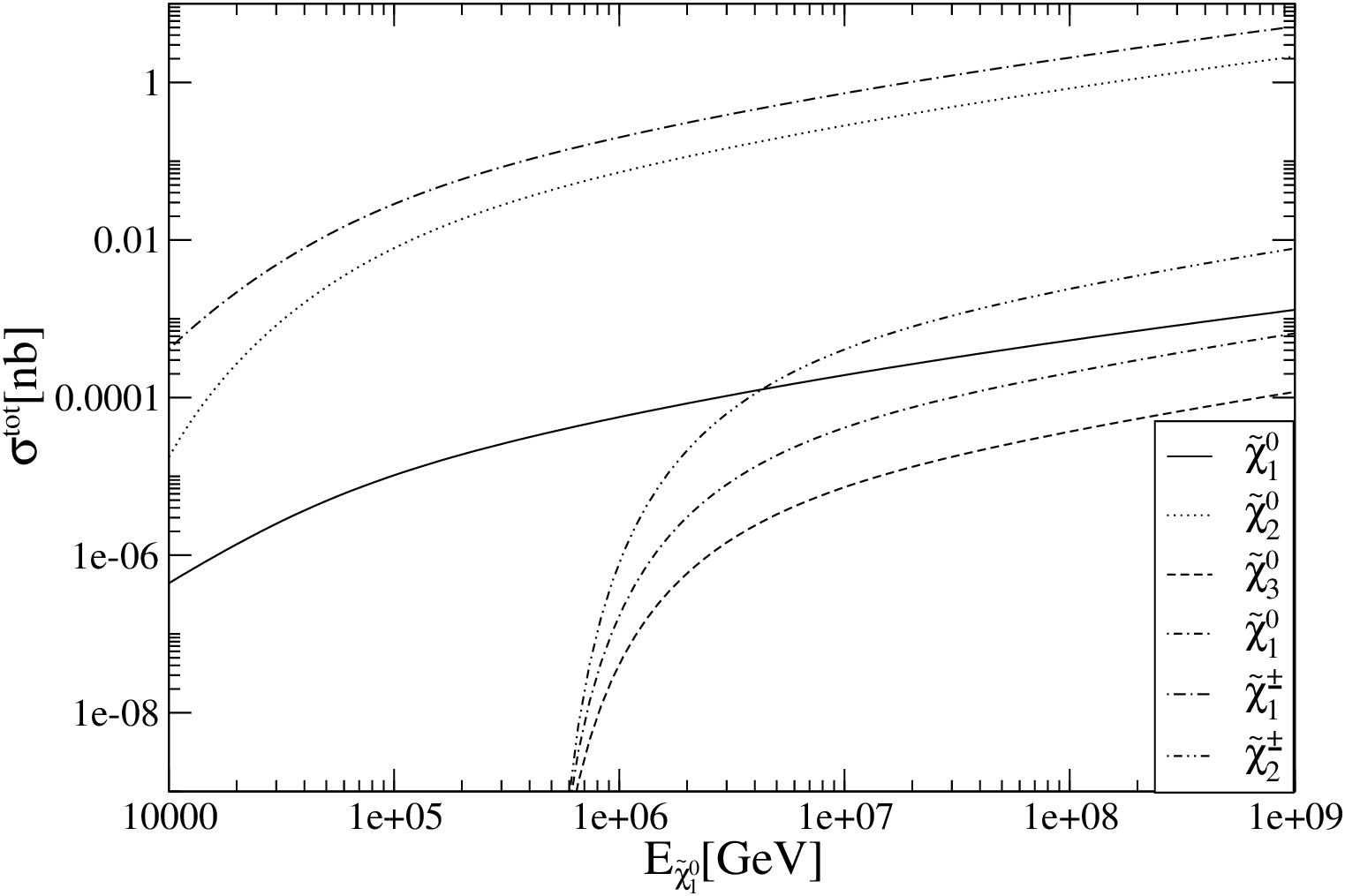}
\caption{\label{fig6}Separate contributions from the four neutralinos and two
  charginos to the total $t-$channel cross section in scenario H2 of
  Table~\ref{tab3}.}
\end{center}
\end{figure}

Scenarios with higgsino--like $\lsp$ have higgsino mass parameter $|\mu|$
significantly smaller than the $SU(2)_L$ and $U(1)_Y$ gaugino mass parameters
$|M_2|$ and $|M_1|$. As a result, the second neutralino $\tilde \chi_2^0$ and
lighter chargino $\tilde \chi_1^\pm$ will also be higgsino--like, but $\tilde
\chi_3^0, \, \tilde \chi_4^0$ and $\tilde \chi_2^\pm$ will be gaugino--like.
Since gauge bosons can only couple to two higgsinos (or, in case of $W^\pm$,
to two $SU(2)_L$ gauginos), the production of these heavier neutralino and
chargino states is strongly suppressed, as shown in Fig.~\ref{fig6}. This
figure shows that the neutral current reaction with $\lsp$ in the final state
is also suppressed. This is due to the fact that $|N_{13}| \simeq |N_{14}|$
for higgsino--like LSP, leading to a strong cancellation of the $Z \lsp \lsp$
couplings in Eqs.(\ref{ncoup}). The total neutral and charged current
contributions are therefore dominated by $\tilde \chi_2^0$ and $\tilde
\chi_1^\pm$ production, respectively. The $\tilde \chi_1^\pm$ cross section is
about two times larger, since $\tilde \chi_2^0$, $\tilde \chi_1^+$ and $\tilde
\chi_1^-$ contribute about equally. All three states are essentially $SU(2)$
doublets, and are produced with nearly pure vector coupling, leading to nearly
identical cross sections for $\tilde \chi_1^\pm$ production on a quark or
antiquark, see Eqs.(\ref{tm1}), (\ref{tm2}).\footnote{In fact, after summing
  over quark and antiquark contributions, at high energies the cross sections
  for $\tilde \chi_1^+$ and $\tilde \chi_1^-$ production are nearly the same
  even if left-- and right--handed couplings $C^{(L,R)}_{11}$ of
  Eqs.(\ref{ccoup}) differ from each other. The reason is that dominant
  contributions come from small values of $x$, where the distribution
  functions for quarks and antiquarks are essentially identical. For the same
  reason the cross section for the scattering of neutrinos and antineutrino
  becomes identical at high energies \cite{GQRS96,GQRS98}.}

\vspace*{10mm}

We now turn to a discussion of the $y-$distribution of the $t-$channel
contribution. The scaling variable $y$ is related to the scattering angle via
\begin{equation} \label{ey}
y = \left( 1 - \cth^* \right) \frac {\hat s - m^2_{\tilde \chi_{\rm out}} }
{2 \hat s}\, .
\end{equation}
The cross section differential in $y$ can therefore be calculated from
Eq.(\ref{ttot}) by a substitution of variables, and inverting the order of
integration:
\eqa \label{dsdy}
\frac {d\sigma_t} {dy} &=& \sum_q \int_{x_-}^1 dx q\!  \kll
 x,Q^2 \klr \frac {|\mathcal{M}|^2} {16\pi (\hat{s} - m^2_{\tilde \chi_1^0}) }
 \, .
\eqe
Eqs.(\ref{ey}) and (\ref{cmax}) imply that for given $x$,
\begin{equation} \label{ylim}
y \in \left[ \frac {Q^2_{\rm min}} {2 x E_{\rm in} m_N}, \ 1 - \frac
  {m^2_{\tilde \chi_{\rm out}}} {\hat s} \right]\, .
\end{equation}
Using $\hat s = m^2_{\tilde \chi_1^0} + 2 x E_{\rm in} m_N$, this gives for
the lower bound on the $x-$integration in Eq.(\ref{dsdy}):
\begin{equation} \label{xm}
x_- = \frac {1} {2 E_{\rm in} m_N} \, \max \left( \frac {Q^2_{\rm min}} {y}, \
\frac {m^2_{\tilde \chi_{\rm out}}} {1-y} - m^2_{\lsp} \right) \, .
\end{equation}
Since $Q^2_{\rm min} = 1$ GeV$^2 \ll m^2_{\lsp}$, for most values of $y$ the
lower bound on $x$ is determined by the second term in Eq.(\ref{xm}); the
restriction on $Q^2$ determines $x_-$ only for very small values of $y$.

\vspace*{10mm}
\begin{figure}[htb] 
\begin{center}
\includegraphics[width=12cm]{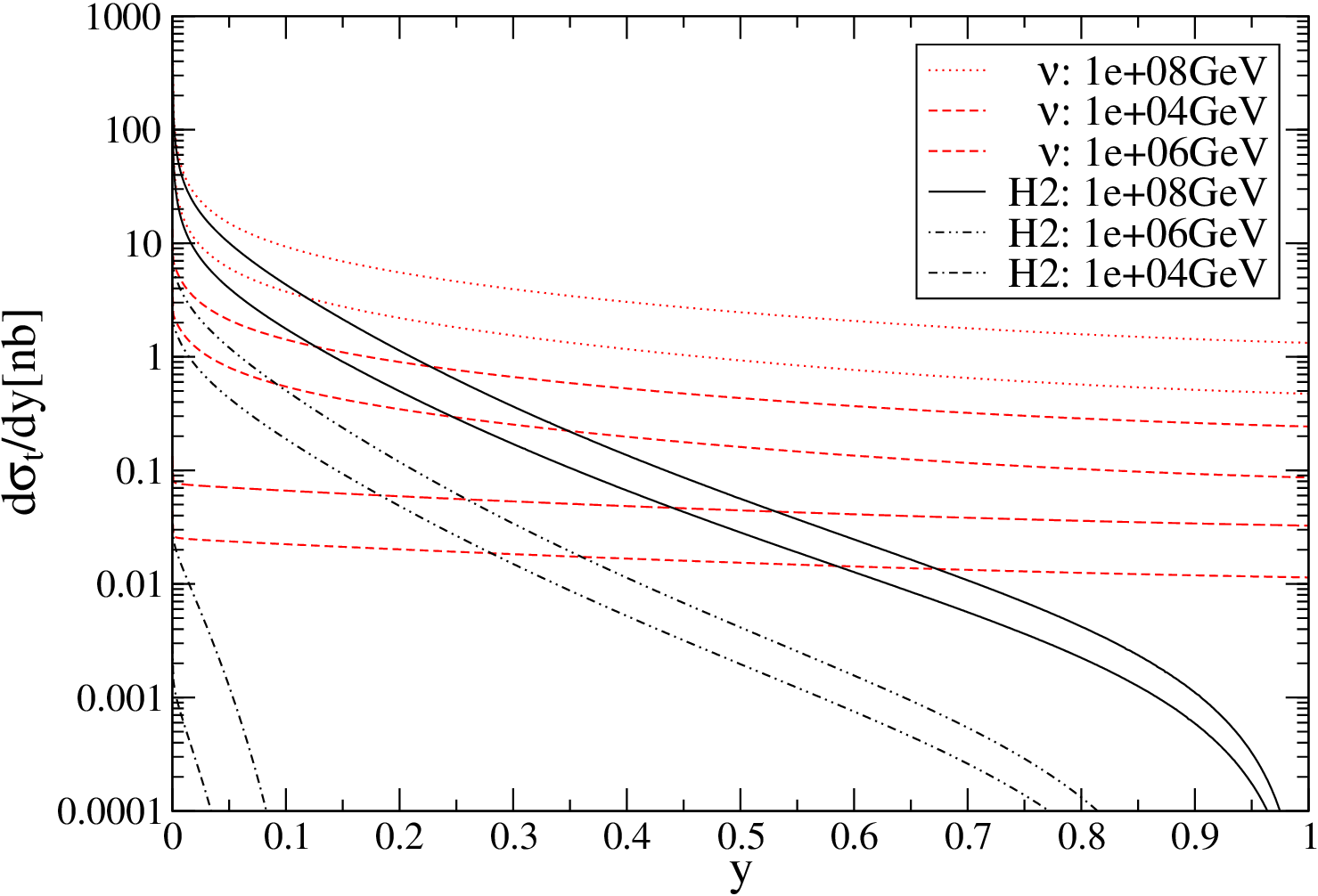}
\caption{\label{fig7}Differential neutral and charged current $t-$channel
  cross sections $d \sigma / dy$ for LSP--nucleon scattering in scenario H2
  (black curves), and for neutrino--nucleon scattering (red or grey curves),
  for three different energies. The charged current contributions are given by
  the upper curve for a fixed energy.}
\end{center}
\end{figure} 

Fig.~\ref{fig7} shows the differential cross section $d \sigma / d y$ for
scenario H2 and three different energies. The charged and neutral current
contributions are shown separately. They have very similar shapes at all
energies. This is not surprising, since in the pure higgsino limit, both
contributions are from vector--like couplings with equal strength. Moreover,
Table~\ref{tab3} shows that the masses of $\tilde \chi_2^0$ and $\tilde
\chi_1^\pm$ are very close to each other; we saw in Fig.~\ref{fig6} that the
production of these two particles dominates the total $t-$channel
contribution.

Fig.~\ref{fig7} also shows the differential cross section for
neutrino--nucleon scattering at the same energies. While neutrino and
neutralino cross sections are quite similar for $y \ll 1$, the latter fall
much more quickly with increasing $y$. This is largely due to the lower bound
(\ref{xm}) on the integral over $x$ in Eq.(\ref{dsdy}), which increases
quickly with increasing $y$. Notice that in the region of small $y$, which
dominates the total cross section, $x_-$ is given by the {\em difference}
between the masses of the incoming and outgoing $\tilde \chi$ particles. Table
~\ref{tab3} shows that this difference is smaller for scenario H2, which has
much larger masses for the higgsino--like states. This explains why the
difference between the total cross sections in scenarios H1 and H2 depicted in
Fig.~\ref{fig5} is smaller than that between the cross section for H1 and the
neutrino cross section. Scenario H2 still gives a smaller cross section since
the mass terms enter the squared matrix elements (\ref{tm1}), (\ref{tm2}) and
(\ref{tm3}) with negative sign. The same effect also explains why the ratio of
charged and neutral current contributions falls slightly with increasing $y$:
the fact that $\tilde \chi_1^\pm$ is slightly lighter than $\tilde \chi_2^0$
is more important at small $y$. In case of neutrinos, $y \rightarrow 1$
remains possible at very small $x$. However, increasing $y$ also means
increasing the squared four--momentum exchange $|\hat t|$ for fixed $\hat s$;
this explains the decrease of the neutrino cross sections.

Up to this point we have not considered the interference between the $s-$ and
$t-$channel contributions. This is suppressed by two effects. First, it can
only occur for outgoing $\lsp$ state; we saw in Fig.~\ref{fig6} that this
final state does not contribute much to the total $t-$channel cross section.
Secondly, the interference between the $t-$channel gauge boson and $s-$channel
squark propagators vanishes at $\hat s = m^2_{\tilde q}$, where the latter is
largest in absolute size, but purely imaginary. Interference can thus only
occur for off--shell squark exchange, and is hence of higher order in
perturbation theory. These interference contributions can therefore safely be
neglected.

\section{Solution of the transport equation}

We are now ready to discuss the propagation of UHE neutralino LSPs through
matter. This is described by so--called transport equations, which have also
been used to describe neutrino propagation through the Earth
\cite{iter,reno1,rr}. We will again treat $s-$ and $t-$channel dominated
scenarios separately; we saw in the previous Section that this corresponds to
scenarios with bino-- and higgsino--like LSP, respectively.

\subsection{Transport equation for $s-$channel scattering}

The quantity we wish to compute is the differential flux $F_{\lsp}$ as
function of the LSP energy $E$ and matter depth $X$, where the latter quantity
is customarily given as a column depth, measured in g/cm$^2$ or, in natural
units, in GeV$^3$; for the Earth, $X \in [0, \, 2.398 \cdot 10^6$ GeV$^3]$
\cite{GQRS96}.\footnote{An LSP exiting the Earth at angle $\theta$ relative to
  the vertical direction will have transversed a column depth $X = X_{\rm max}
  \cth$.} A UHE LSP interacting with matter at rest can only lose energy. The
interactions of LSPs with energy $E$ will therefore always reduce
$F_{\lsp}(E)$. The size of this effect is determined by the interaction length
$\lambda_{\lsp}$, which is given by
\begin{equation} \label{el}
\lambda_{\lsp}(E)^{-1} = N_A \sigma^{\rm tot}_{\lsp N}(E)\, ,
\end{equation}
where $N_A=6.022\times 10^{23}\mbox{~g}^{-1}$ is Avogadro's number, and in the
given scenario $\sigma^{\rm tot}_{\lsp N} \simeq \sigma^{\rm tot}_s$ of
Eq.(\ref{es}). On the other hand, $F_{\lsp}(E)$ can be {\em increased} by the
interactions of LSPs with energy $E_y \equiv E/(1-y) > E$ losing a fraction
$y$ of their energy. These considerations lead to the transport equation
\eqa \label{ts}
\frac {\partial F_{\lsp}(E,X)} {\partial X} &=& -\frac {F_{\lsp}(E,X)}
{\lambda_{\lsp}(E)} + \frac {1} {\lambda_{\lsp}(E)}
 \int_0^{y_{\rm max}} \frac {dy} {1-y} K_s(E,y) F_{\lsp}(E_y,X) \, ,
\eqe
where the kernel is determined by the differential cross section given in
Eq.(\ref{diffxs}):
\eqa \label{ks}
K_s(E,y) &=& \frac {1} {\sigma^{\rm tot}_s(E)} \frac {d\sigma_s(E_y)} {dy} \, ,
\eqe
and $y_{\rm max}$ has been given in Eq.(\ref{ymax}). Note that Eq.(\ref{ts})
assumes collinear kinematics, where the produced LSP goes in the same
direction as the original one. This is justified for ultra--relativistic LSPs,
whose scattering angle is $\theta \lsim m_{\tilde q}/E_{\rm in} \lsim 10^{-3}$
for energies of interest.

An important property of $\lsp-$nucleon scattering is that it always produces
another $\lsp$ in the final state. The total flux, 
\begin{equation} \label{ephi}
\Phi_{\lsp} = \int_{m_{\lsp}}^{E_{\rm max}} F_{\lsp}(E,X) dE\, ,
\end{equation}
must therefore remain constant, independent of $X$; here $E_{\rm max}$ is the
maximal $\lsp$ energy, beyond which the incident LSP flux vanishes. This is
reflected in the fact that integrating the right--hand side of Eq.(\ref{ts})
over the energy gives zero, i.e.~$d \Phi_{\lsp} / dX = 0$.\footnote{To see
  this, one re--writes the double integral over $E$ and $y$ into an integral
  over $E_y$ and $y$ and uses the definition (\ref{el}) of the $\lsp$
  interaction length.} This allows an important check of the procedure used to
solve the transport equation.

The standard method to solve the analogous transport equation for neutrinos is
based on an iteration \cite{iter,reno1,rr}. It uses an effective interaction
length $\Lambda_{\lsp}$ defined by
\begin{equation} \label{ei1}
F_{\lsp}(E,X) = F_{\lsp}(E,0) \exp \left[ - \frac {X} {\Lambda_{\lsp}(E,X)}
\right]
\, ,
\end{equation}
which in turn leads to the definition of a ``$Z-$factor'':
\begin{equation} \label{ei2}
\Lambda_{\lsp}(E,X) = \frac {\lambda_{\lsp}(E)} { 1 - Z_{\lsp}(E,X) } \, .
\end{equation}
The transport equation can then be re--written as an integral equation for
$Z_{\lsp}$. This equation can be solved iteratively, starting from the $0-$th
order ansatz $Z^{(0)}_{\lsp} = 0$. This treatment was first introduced in
ref.\cite{iter} to describe the propagation of electron and muon
neutrinos. In this case the iteration converges fast, so that the first
non--trivial solution $Z^{(1)}_{\nu_\mu}$ deviates by at most 4\% from the
final answer. The same first--order solution was then used in
refs.\cite{reno1,rr} to describe $\nu_\tau$ propagation. We therefore tried to
use it for LSP propagation as well.

Unfortunately we found that the first--order solution badly violates flux
conservation in our case. A close reading of ref.\cite{iter} shows that this
is actually not very surprising. To begin with, the $0-$th order solution
evidently ignores regeneration completely. Flux conservation requires that the
differential flux must increase for some range of energies; Eqs.(\ref{ei1})
and (\ref{ei2}) show that this is possible only if the $Z-$factor exceeds
unity, which means that the effective interaction length $\Lambda$ must become
negative. This did not happen in ref.\cite{iter}, since charged current
reactions of electron or muon neutrinos effectively lead to a loss of this
neutrino\footnote{In the energy range of interest, muons will come to rest
  before decaying. The $\nu_\mu$ produced in this decay has such a low energy
  that it is effectively invisible to neutrino telescopes.}, thereby reducing
the total neutrino flux. Flux conservation was therefore of no concern in
ref.\cite{iter}.  Moreover, the neutral current reaction, whose regeneration
effect was included in this treatment, has $d \sigma / d y$ quite strongly
peaked at low $y$, see Fig.~\ref{fig7}. According to ref.\cite{iter}, this
accelerates the convergence of the iteration. However, in our case $d \sigma_s
/ d y$ is flat, see Eq.(\ref{diffxs}). It is therefore not surprising that
this algorithm does not work very well in our case.\footnote{We suspect that
  the first order solution of this algorithm also leads to sizable errors when
  applied to $\nu_\tau$ propagation. Since $\tau$ leptons in the relevant
  range of energies decay before they interact, the total $\nu_\tau$ flux
  should be conserved. Moreover, since only a relatively small fraction of the
  energy of the original $\nu_\tau$ goes into the $\nu_\tau$ produced in
  $\tau$ decay, the effective $y$ distribution, computed as in Eq.(\ref{kt})
  below, should extend to quite large values of $y$.}

We therefore use a straightforward numerical solution of the transport
equation (\ref{ts}), based on the first order Taylor expansion:
\eqa \label{taylor}
F_{\lsp}(E,X+dX) = F_{\lsp}(E,X) + dX \frac{\partial
  F_{\lsp}(E,X)} {\partial X} + \cdots  \, .
\eqe
We found that a more sophisticated algorithm, e.g. the Runge--Kutta method,
does not offer much of an advantage in terms of accuracy achieved for a fixed
CPU time spent. We parameterize $F_{\lsp}(E)$ for given $X$ as cubic spine.
The one--dimensional integral appearing in Eq.(\ref{ts}) can then be evaluated
using, e.g., the Simpson method.

Since the transport equation is a first order differential equation, we have
to specify the boundary condition $F_{\lsp}(E,0)$. We use three different
ans\"atze: 
\eqa \label{ef}
\textrm{Spectrum \ 1:} \hspace{0.1cm} F_{\lsp}^{0}(E, 0) &=& N_1 E^{-2}
F_{\rm cut}(E) \, ; \nonumber \\
\textrm{Spectrum \ 2:} \hspace{0.1cm} F_{\lsp}^{0}(E, 0) &=& N_2 E^{-1.5}
  F_{\rm  cut}(E) \, ; \nonumber \\
\textrm{Spectrum \ 3:} \hspace{0.1cm} F_{\lsp}^{0}(E, 0) &=& N_3 E^{-1}
  \frac{1}{\kll 1 + \frac {E} {10^8\ {\rm GeV}} \klr^2} F_{\rm cut}(E)
\, ,
\eqe
where the $N_i$ are arbitrary normalization factors. The first two ans\"atze
would yield an infinite total energy stored in LSPs, had we not introduced a
cut--off function
\begin{equation} \label{ecut}
F_{\rm cut}(E) = 1- \kll \frac {E} {E_{\rm cut}} \klr^4 \, ,
\end{equation}
with $F_{\rm cut} (E) =0$ for $E > E_{\rm cut}$.  In top--down scenarios,
$E_{\rm cut}$ should be $\sim M_X/2$. In our numerical examples we instead use
$E_{\rm cut} = 10^9$ GeV. We will see below that LSPs with initial energy $E
>10^9$ GeV will interact very soon, i.e.~the flux at $E > 10^9$ GeV will
quickly become negligible. Due to the shorter interaction length, choosing
$E_{\rm cut} > 10^9$ GeV would greatly increase the numerical cost of the
solution. On the other hand, this relatively small value of $E_{\rm cut} \ll
M_X$ is justified if neutralinos with $E > E_{\rm cut}$ contribute negligibly
to the total LSP flux impinging on Earth, which is often the case \cite{bad}.

\vspace*{6mm}
\begin{figure}[h!] 
\begin{center}
\includegraphics[width=8cm]{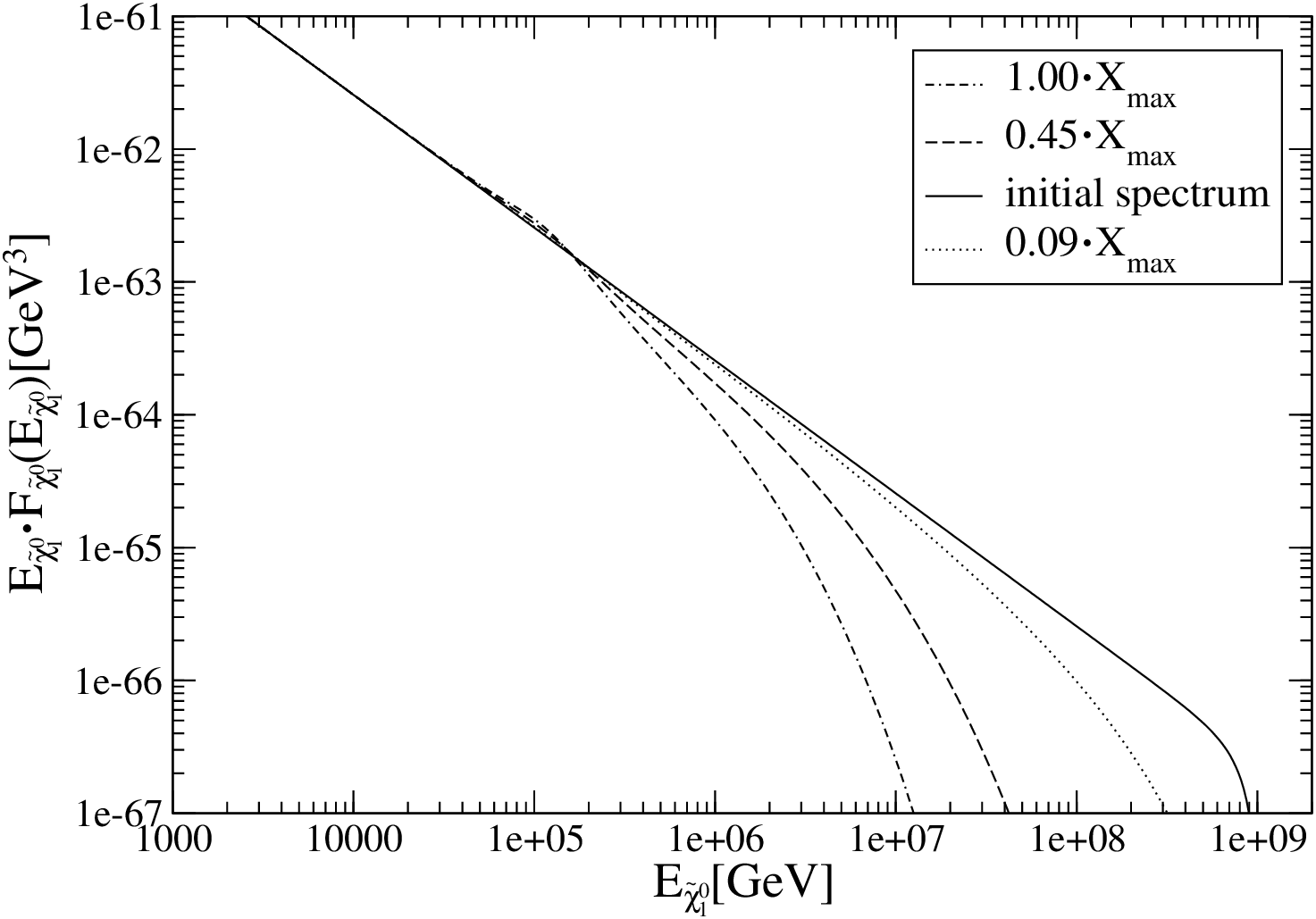} \hspace*{3mm}
\includegraphics[width=8cm]{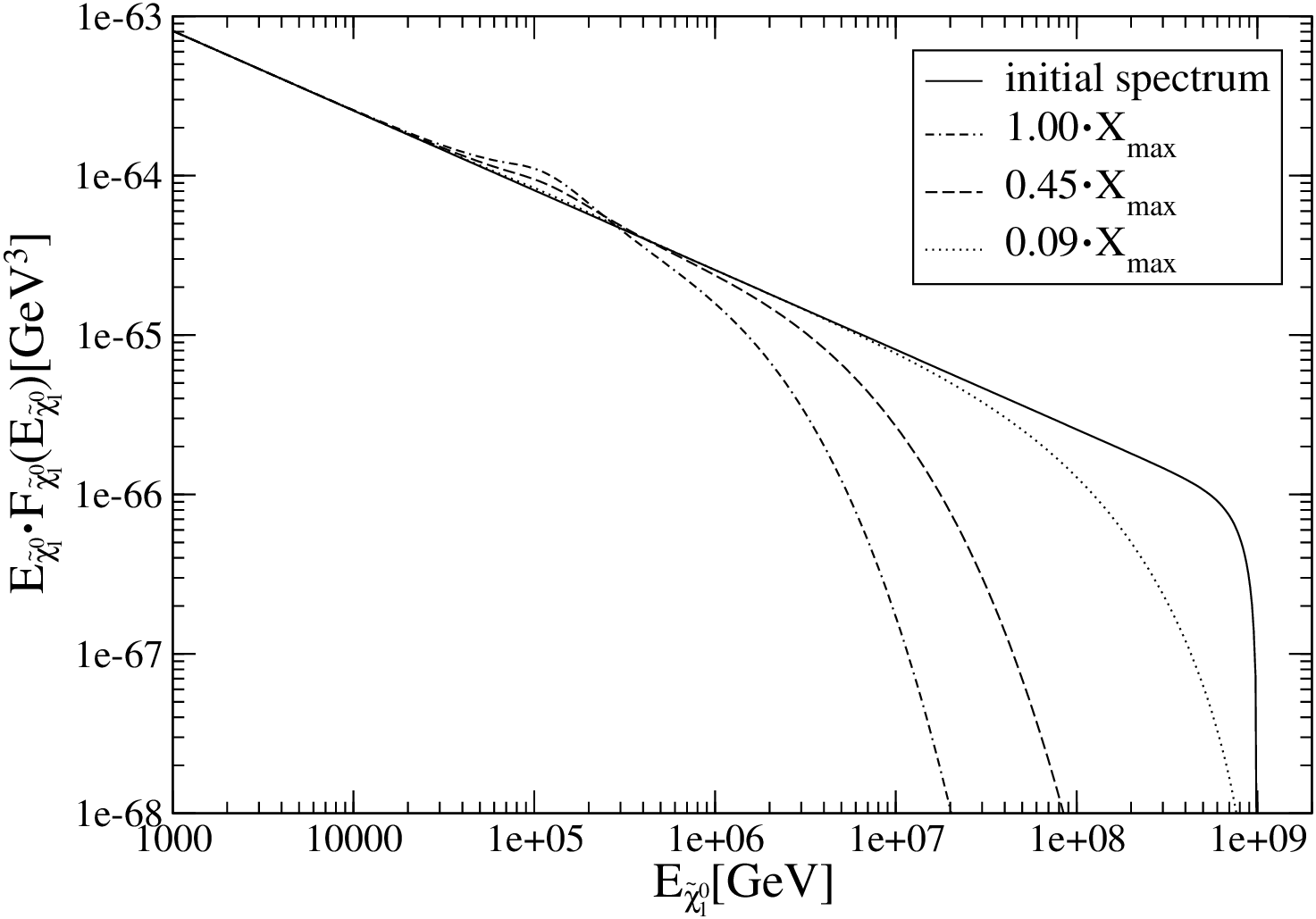} 

\vspace*{9mm}
\includegraphics[width=8cm]{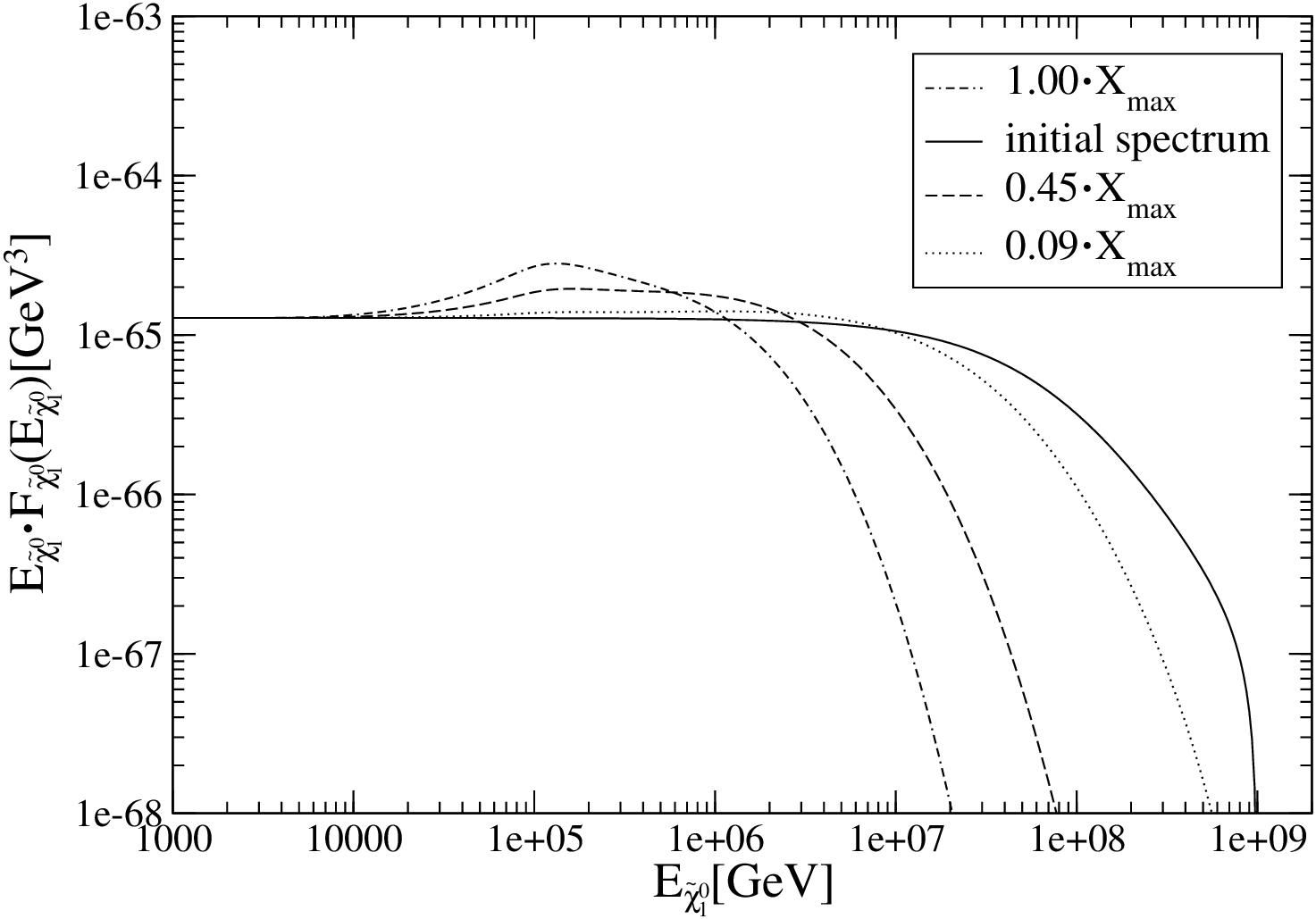}
\caption{\label{fig8}Modification of our three LSP spectra due to propagation
  through the Earth for scenario D1 with bino--like LSP. The solid curve shows
  the initial spectra (in arbitrary units), and the dotted, dot--dashed and
  dashed curves show the flux after a column depth of $0.09 X_{\rm max}, \,
  0.45 X_{\rm max}$ and $X_{\rm max} = 2.4 \cdot 10^6$ GeV$^3$. The top--left,
  top--right and bottom frames are for Spectrum 1, Spectrum 2 and Spectrum 3
  of Eqs.(\ref{ef}), respectively.}
\end{center}
\end{figure}

In this Subsection we focus on the first scenario of Table~\ref{tab1}. We saw
in Fig.~\ref{fig2} that this leads to the largest cross section, and hence the
largest propagation effects. Results for the three initial fluxes of
Eq.(\ref{ef}) are shown in Fig.~\ref{fig8}. The initial spectrum should get
reduced significantly once the traversed column depth exceeds the interactions
length; note that according to Eq.(\ref{el}), $\sigma^{\rm tot}_{\lsp N} = 1$
nb corresponds to an interaction length $\lambda_{\lsp} \simeq 0.36 \cdot
10^6$ GeV$^3 \simeq 0.15 X_{\rm max}$. Fig.~\ref{fig2} shows that in scenario
D1 this cross section is reached at $E_{\rm in} \simeq 10^7$ GeV. We
see in the two upper frames of Fig.~\ref{fig8} that the suppression of the LSP
spectrum indeed begins to be noticeable around that energy for $X = 0.09 X_{\rm
  max}$.

In the lower frame the suppression starts at somewhat higher energy for this
small column depth, since for this very hard initial spectrum the regeneration
effect is maximized. This effect creates a bump in the spectrum. The peak of
the bump is at an energy such that the interaction length is slightly larger
than the traversed column depth; since the cross section rises with energy,
the bump shifts to lower energies with increasing column depth. Conservation
of the total flux (\ref{ef}) implies that this bump is more noticeable for a
harder initial spectrum. An analogous argument explains why the energy where
the spectrum begins to be suppressed becomes smaller with increasing column
depth. This can also be understood from the simple observation that increasing
column depth increases the probability for interactions between LSPs and
matter.

Fig.~\ref{fig2} shows that at high energies, the total $s-$channel cross
section for bino--like LSP is roughly proportional to $(1/m^2_{\tilde q})
(E_{\rm in}/m_{\tilde q})^{0.4}$. According to Table~\ref{tab1}, squarks are
about 2.7 times heavier in scenario D3 than in scenario D1. The onset of the
suppression of the initial LSP spectrum, and the location of the bump produced
by regeneration, will thus occur at about 12 times higher energy than in
scenario D1.

As discussed at the beginning of this Subsection, the total LSP flux
(\ref{ephi}) must be conserved. We checked that our numerical solution
satisfies this constraint very well, with maximal deviation of less than 0.1\%
even for the hardest spectrum and largest column depth. In contrast, for the
hardest incident LSP spectrum the prediction of the first--order iterative
solution can violate flux conservation by a factor of two or more.

\subsection{Transport equation for $t-$channel scattering}

We consider the transport equation for $t-$channel scattering only for a
scenario with higgsino--dominated LSP; Fig.~\ref{fig4} showed that $t-$channel
contributions are negligible for a bino--like neutralino. As in case of
neutrinos \cite{iter,reno1,rr}, there are both charged and neutral current
contributions. We saw in Fig.~\ref{fig6} that these predominantly lead to the
production of $\tilde \chi_1^\pm$ and $\tilde \chi_2^0$, respectively. Here we
will ignore the contribution from all other final states; this approximation
is good at the few \% level. 

In case of $\nu_\tau$ propagation \cite{reno1,rr} one usually writes two
coupled transport equations, for $\nu_\tau$ itself and for the $\tau$ leptons
produced in $\nu_\tau$ charged current reactions. Correspondingly we would
need three coupled transport equations in our case, describing the spectra of
$\lsp, \, \tilde \chi_1^\pm$ and $\tilde \chi_2^0$. However, these heavier
produced particles decay well before they lose a significant fraction of their
energy through interactions. This is true even for $\tau$ leptons of the
relevant energy; the $\tilde \chi_1^\pm$ and $\tilde \chi_2^0$ states in the
scenarios of interest have much shorter lifetimes [by a factor of order
$(m_{\tilde \chi_{\rm out}} - m_{\lsp})^5/m_\tau^5$], and significantly
shorter decay lengths for a given lifetime (due to their larger masses,
i.e.~smaller $\gamma$ factors). $t-$channel scattering can thus be treated
through an integration kernel in the transport equation which is computed from
a convolution of the $\tilde \chi_{\rm out}$ production and decay spectra. We
treat charged and neutral current contributions separately:
\eqa \label{tt}
\frac {\partial F_{\lsp}(E,X)} {\partial X} &=& -\frac {F_{\lsp}(E,X)}
{\lambda_{\lsp}(E)} + \frac{1} {\lambda_{\lsp}(E)}
 \int_0^1 \frac {dy} {1-y} K_{\lsp}^{NC}(E,y) F_{\lsp}(E_y,X)  \nonumber \\
 &&+\frac {1} {\lambda_{\lsp}(E)} \int_0^1 \frac {dy} {1-y} K_{\lsp}^{CC}(E,y)
 F_{\lsp}(E_y,X) \, ,
\eqe
where the interaction length $\lambda_{\lsp}$ has been given by Eq.(\ref{el}).
The structure of the equation is very similar to Eq.(\ref{ts}) describing
$s-$channel scattering. The integration kernels are most easily written as
convolutions in the variable $z = 1 - y = E_{\rm out} / E_{\rm in}$:
\eqa \label{kt}
K_{\lsp}^{NC,CC}(E,y) &=& \frac {1} {\sigma_t(E)}
\int_{z_{1,{\rm min}}}^{z_{1,{\rm max}}} d\!z_1 \int_{z_{2,{\rm
      min}}}^{z_{2,{\rm max}}} d\!z_2 \delta\!  \kll z-z_1z_2\klr  
\frac{d\sigma^{NC,CC} (E_y,z_1)}{dz_1} \left.
\frac{1}{\Gamma}\frac{d\Gamma(z_1 E_y,z_2)}{dz_2} \right|_{y = 1-z}
\nonumber \\
&=&  \frac {1} {\sigma_t(E)} \int_z^{z_{1,{\rm max}}} \frac{d\!z_1}{z_1}
\frac{d\sigma^{NC,CC} (E_y,z_1)}{dz_1} \left.
\frac {1} {\Gamma} \frac {d\Gamma(z_1E_y, \frac{z}{z_1})}{dz_2}
\theta(z-z_{\rm min}) \theta(z_{\rm max} - z) \right|_{y = 1-z} \, .
\nonumber \\ &&
\eqe
The limits $z_{1,{\rm min, \, max}}$ for the outer integration in the first
line of Eq.(\ref{kt}) follow from Eq.(\ref{ylim}) with $x \rightarrow 1$,
i.e.~$\hat s \rightarrow s = 2 E_y m_N + m_{\lsp}^2$. The limits for the inner
integration follow from $\tilde \chi_{\rm out}$ decay kinematics described
below. The $\theta-$functions in the second line of Eq.(\ref{kt}) ensure that
$y$ lies within the kinematical limits, with $z_{\rm min} = z_{1,{\rm min}}
z_{2,{\rm min}}$ and $z_{\rm max} = z_{1,{\rm max}} z_{2,{\rm max}}$. Note
that both integration kernels in Eq.(\ref{tt}) are normalized to the {\em
  total} $\lsp-$nucleon scattering cross section, which is here approximated
by the $t-$channel contribution given by Eq.(\ref{ttot}). Finally,
Eq.(\ref{tt}) again assumes collinear kinematics, where the LSP produced in
$\tilde \chi_{\rm out}$ decay goes into the same direction as the original
LSP.

The missing piece in Eq.(\ref{kt}) is the differential decay spectrum of the
produced $\tilde \chi_{\rm out}$. Due to the small $\tilde \chi_{\rm out}-
\lsp$ mass difference for higgsino--like LSP, see Table~\ref{tab3}, we only
need to consider three--body decays, $\tilde \chi_{\rm out} \rightarrow \lsp
f_1 f_2$, where $f_1, \, f_2$ are two SM (anti)fermions whose masses we
neglect. In the $\tilde \chi_{\rm out}$ rest frame we then have:
\eqa \label{gam1}
\frac{d\Gamma}{dE^*_{\lsp}}= \frac {1} {8m_{\tilde \chi_{\rm out}} \kll 2 \pi
  \klr^3} \int_{E_{f,{\rm min}}^*}^{E_{f,{\rm max}}^*} dE_f^* |{\cal M}|^2 \, ,
\eqe
where $E_f^*$ is the energy of one of the two massless (anti)fermions (the
energy of the other being determined by energy conservation). The integration
limits in Eq.(\ref{gam1}) are given by
\eqa \label{eflim}
E_{f \stackrel {\rm min} {\rm \scriptscriptstyle{max} }}^* = \frac {2
  m_{\tilde \chi_{\rm out}} E^*_{\lsp} -
    m_{\lsp}^2 - m_{\tilde \chi_{\rm out}}^2 } { 2 E_{\lsp}^* - 2 m_{\tilde
      \chi_{\rm out}} \pm 2 \sqrt{E_{\lsp}^{*2}-m_{\lsp}^2} } \, .
\eqe
The total $\tilde \chi_{\rm out}$ decay width $\Gamma$ appearing in
Eq.(\ref{kt}) can be obtained by integrating Eq.(\ref{gam1}) over
$E^*_{\lsp}$, the lower and upper integration limit being given by $m_{\lsp}$
and $(m_{\lsp}^2 + m^2_{\tilde \chi_{\rm out}}) / (2 m_{\tilde \chi_{\rm
    out}})$, respectively.

In order to boost into the frame where $\tilde \chi_{\rm out}$ has energy
energy $E_{\tilde \chi_{\rm out}}$, we have to know the angular distribution of
the produced $\lsp$ in the $\tilde \chi_{\rm out}$ rest frame relative to the
$\tilde \chi_{\rm out}$ flight direction. Here we assume an isotropic
distribution, which is appropriate for an unpolarized $\tilde \chi_{\rm out}$,
and also if $\tilde \chi_{\rm out}$ decays via a pure vector coupling; we saw
in Subsec.~2.2 that the relevant couplings are indeed almost pure vector
couplings in scenarios with higgsino--like LSP. This yields
\eqa \label{gam2}
\frac {d\Gamma} {dE_{\lsp}} = \int_{E^*_{\lsp,{\rm min}}}^{E^*_{\lsp,{\rm
      max}}} d\!E_{\lsp}^* \frac {m_{\tilde \chi_{\rm out}}} {E_{\tilde
    \chi_{\rm out}}} \frac {1} {\beta \sqrt{E_{\lsp}^{*2} - m_{\lsp}^2}} \frac
{1} {16m_{\tilde\chi_{\rm out}} \kll2 \pi \klr^3} \int_{E_{f,{\rm min}}^*}^{E_{f,{\rm max}}^*}
dE_f^* |\mathcal{M}|^2 \, .
\eqe
The limits for the inner integration in Eq.(\ref{gam2}) have been given in
Eq.(\ref{eflim}), and the limits for the outer integration are:
\eqa \label{estlim}
E^*_{\lsp,{\rm min}} &=& \gamma \left( E_{\lsp} - \beta \sqrt{E^2_{\lsp} -
    m^2_{\lsp}} \right)\, , \nonumber \\
E^*_{\lsp,{\rm max}} &=& \frac {m^2_{\tilde \chi_{\rm out}} + m^2_{\lsp}}
{2 m_{\tilde \chi_{\rm out}}} \, .
\eqe
Here, and in Eq.(\ref{gam2}), $\gamma = 1/\sqrt{1-\beta^2} = E_{\tilde
  \chi_{\rm out}} / m_{\tilde \chi_{\rm out}}$. Notice that $E^*_{\lsp,{\rm
  min}}$ reduces to the absolute kinematical minimum of $m_{\lsp}$  for
$E_{\lsp} = \gamma m_{\lsp}$, whereas $E^*_{\rm max}$ is always determined
from the $\tilde \chi_{\rm out}$ decay kinematics, independent of
  $E_{\lsp}$. Finally, in the relevant limit $\gamma \gg 1$ the limits on the
  energy $E_{\lsp}$ of the LSP produced in $\tilde \chi_{\rm out}$ decay are
\eqa \label{elim}
E_{\lsp,{\rm min}} &=& \frac {\gamma \left[ m_{\lsp}^2 + m_{\tilde \chi_{\rm
      out}}^2 - \beta \kll m_{\tilde \chi_{\rm out}}^2 - m_{\lsp}^2 \klr
\right] } {2 m_{\tilde \chi_{\rm out}} } \longrightarrow \gamma \frac
{m^2_{\lsp}} {m_{\tilde \chi_{\rm out}}} \, , \nonumber \\
E_{\lsp,{\rm max}} &=& \frac {\gamma \left[ m_{\lsp}^2 + m_{\tilde \chi_{\rm
      out}}^2 + \beta \kll m_{\tilde \chi_{\rm out}}^2 - m_{\lsp}^2 \klr
\right] } {2 m_{\tilde \chi_{\rm out}} } \longrightarrow \gamma m_{\tilde
      \chi_{\rm out}} = E_{\tilde \chi_{\rm out}}\, .
\eqe
Since the heavier higgsinos $\tilde \chi_{\rm out}$ have very similar masses
as the LSP, see Table~\ref{tab3}, the energy loss in the $\tilde \chi_{\rm
  out}$ decay process cannot be very large, and may be zero. For example, in
scenario H2 the maximal energy loss is 6.7\% in $\tilde \chi_2^0$ decays
produced in neutral current events, and just 2\% in $\tilde \chi_1^\pm$ decays
produced in charged current events. For simplicity we therefore treated these
decays with purely kinematical decay distributions, i.e.~we used a constant
matrix element ${\cal M}$ in Eqs.(\ref{gam1}) and (\ref{gam2}).

\begin{figure}[htb] 
\begin{center}
\includegraphics[width=8cm]{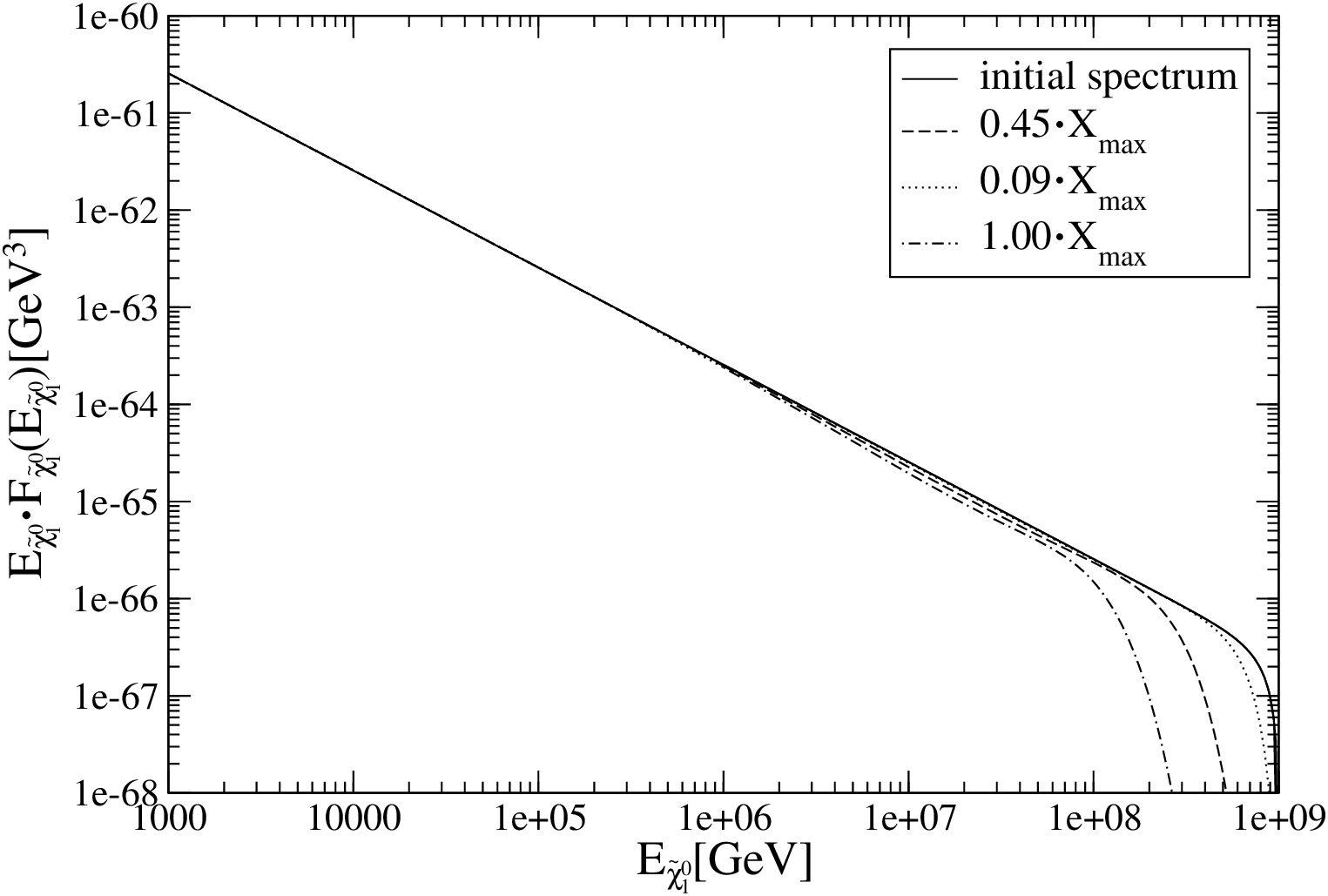} \hspace*{3mm}
\includegraphics[width=8cm]{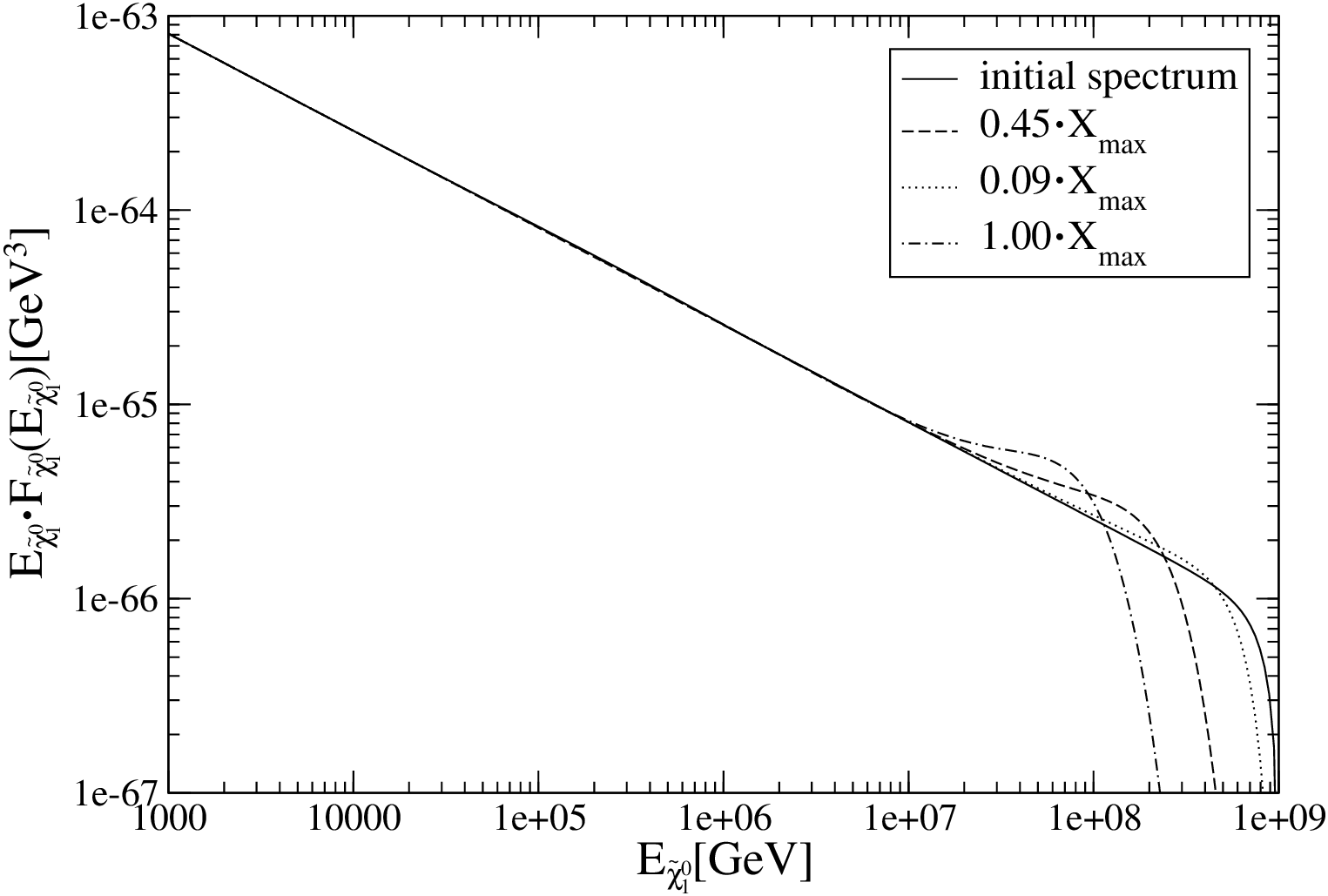}

\vspace*{9mm}
\includegraphics[width=8cm]{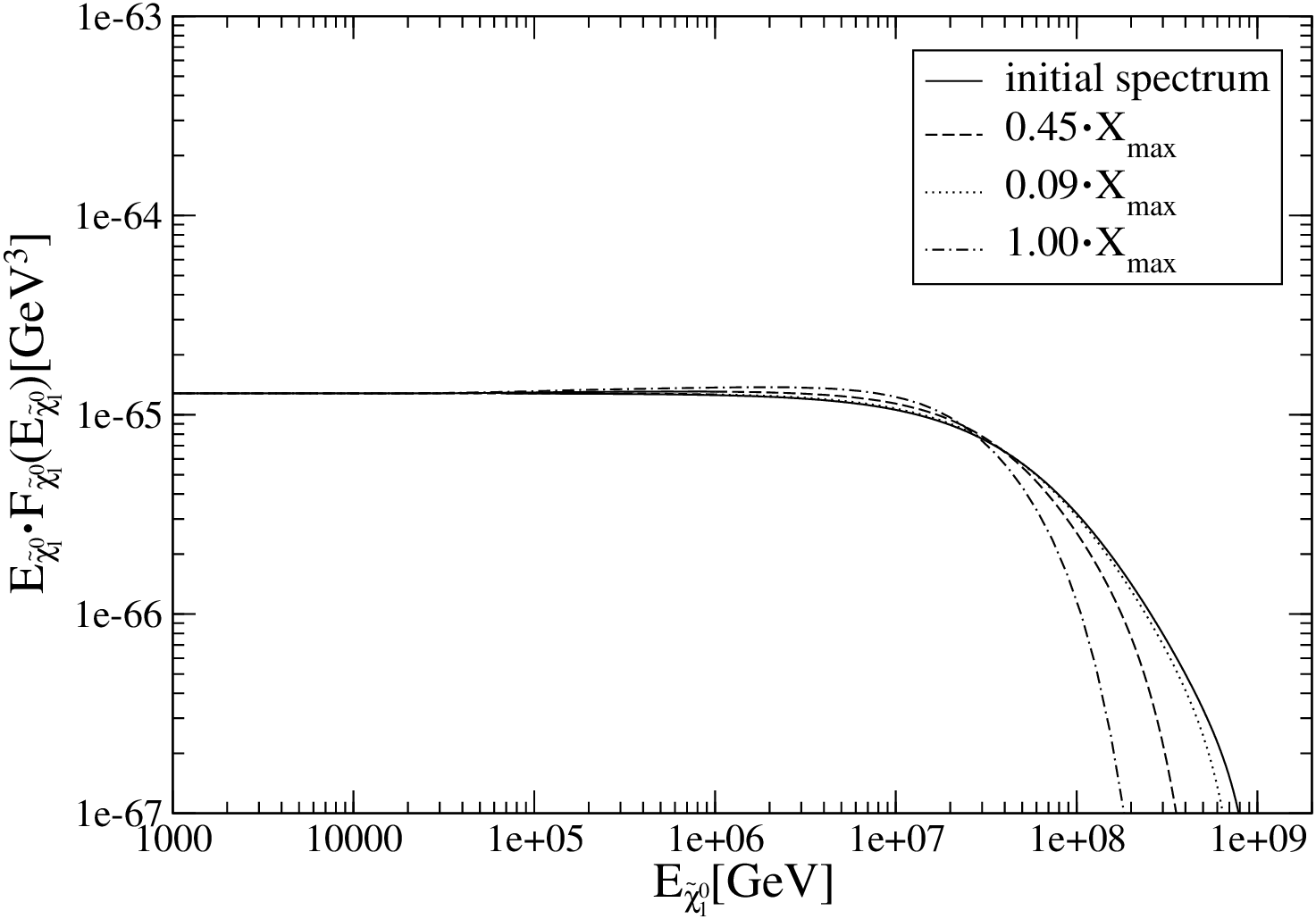}
\caption{\label{fig9}Modification of our three LSP spectra due to propagation
  through the Earth for scenario H2 with higgsino--like LSP. Conventions are
  as in Fig.~\ref{fig8}.}
\end{center}
\end{figure}

Our results for scenario H2 with a higgsino--like LSP are displayed in
Fig.~\ref{fig9}. Although comparison of Figs.~\ref{fig4} and \ref{fig5} shows
that this scenario and the scenario D1 with bino--like LSP lead to similar
total $\lsp-$nucleon scattering cross sections, comparison of Figs.~\ref{fig8}
and \ref{fig9} shows that the spectrum of higgsino--dominated LSPs is modified
much less by propagation effects. The reason is that the $t-$channel exchange
contributions have a far steeper $y-$distribution, cf.~Fig.~\ref{fig7}. The
spectrum will be modified significantly only once a given LSP loses a sizable
fraction of its energy through interactions. The characteristic length for
this is better characterized by $\lambda_{\lsp}^{\rm eff} \equiv
\lambda_{\lsp} / \langle y \rangle$ than by the interaction length
$\lambda_{\lsp}$ itself. Here $\langle y \rangle$ is the average relative
energy loss in an interaction and, for the $t-$channel, subsequent $\tilde
\chi_{\rm out}$ decay; numerically, $\langle y \rangle \simeq 0.5$ for
scenario D1, whereas $\langle y \rangle \simeq 0.07$ for scenario H2.
$\lambda_{\lsp}^{\rm eff}$ therefore differs by about a factor of 10 between
scenarios D1 and H2. As a result, even after traversing the whole Earth in
scenario H2 only the spectrum at $E \gsim 10^8$ GeV is suppressed, whereas in
scenario D1 with bino--like LSP the suppression already starts at $E \sim
10^6$ GeV. The same effect also explains the reduced size of the
``regeneration bump'' in the spectrum of higgsino--like LSPs. Indeed, for the
steepest input spectrum (top--left frame) this feature is no longer visible;
however, Table~\ref{tab5} shows that flux conservation is satisfied also in
this case.\footnote{Due to the steeper $y$ distribution, the first iteration
  of the method developed in ref.\cite{iter} works slightly better in this
  case than for bino--like LSP. However, after traversing the whole Earth we
  still find that up to 40\% of the original LSP flux is ``lost'' if this
  method is used.} We should mention here that due to the additional
integrations required in Eqs.(\ref{kt}) and (\ref{gam2}), this scenario is
numerically much more time consuming than the case of a bino--like LSP.

\vspace*{6mm}

\begin{table}[htb] 
\begin{center}
\begin{tabular}{|c||c|c|c|} 
\hline
& \multicolumn{3}{|c|} {$\Phi_{\lsp}(X) / \Phi_{\lsp}(0)$} \\
$X/X_{\rm max}$ &  Spectrum 1 & Spectrum 2 & Spectrum 3 \\
\hline
0.09 & 1.001 &  1.000  & 1.003 \\
0.45 & 1.001 &  1.001  & 1.013  \\ 
1. & 1.001 &  1.002  & 1.022  \\
\hline
\end{tabular}
\caption{\label{tab5}Test of conservation of the total LSP flux (\ref{ephi})
  for the three incident spectra defined in Eqs.(\ref{ef}) using scenario H2
  with higgsino--like LSP; the LSP spectra are integrated between $10^3$ GeV
  and $E_{\rm cut} = 10^9$ GeV. The column depths are in units of the maximal
  Earth column depth $X_{\rm max} = 2.4 \cdot 10^6$ GeV$^3$.}
\end{center}
\end{table}

\section{Summary and conclusions}

The detection of ultra relativistic cosmic neutralinos $\lsp$ would be a
crucial test of both ``top-down'' models and supersymmetry. This should in
principle be feasible by using the Earth to filter out very energetic
neutrinos, which would otherwise form an unsurmountable background.  In order
to assess the prospects of future or planned experiments such as EUSO
\cite{euso} and OWL \cite{owl} a realistic treatment of the interactions of
neutralinos with ordinary matter is essential. In this paper we provide two
important ingredients of this treatment: an improved calculation of the
LSP--nucleon scattering cross section, and a reliable treatment of propagation
effects by means of transport equations.

The calculation of the most important contributions to the $\lsp-$nucleon
scattering cross sections are presented in Sec.~2. New features include the
calculation of the cross section differential in the energy of the outgoing
neutralino, which is needed for the quantitative description of neutralino
propagation through the Earth, and a first calculation of $t-$channel
scattering processes, which dominate in case of higgsino--like LSP but are
negligible for bino--like LSP. We also calculated the contribution from $\lsp
+ g \rightarrow \tilde t + \bar t$, see the Appendix. This process turns out
to be sub--dominant in the scenarios we studied, but might be important in
special scenarios where the lighter scalar top $\tilde t_1$ is much lighter
than the other squarks. We found that the total cross sections for bino-- and
higgsino--like LSPs are comparable to each other if first and second
generation squark masses are around 400 GeV. For lighter (heavier) squarks the
cross section for bino--like LSPs will be enhanced (suppressed). The cross
section for higgsino--like LSP is much less dependent on the values of various
parameters appearing in the SUSY Lagrangian, as long as the LSP is indeed
higgsino--like; at high energies it is only a factor $\sim 1.5$ below the
cross section for neutrino--nucleon scattering.

However, the shape of the distribution in the energy loss variable $y$ is
quite different in the three cases. Under the assumption that squarks are
lighter than gluinos, a bino--like LSP has a flat $y-$distribution, with
average $\langle y \rangle \simeq 0.5$. In case of neutrinos, the $y$
distribution peaks at $y=0$, but falls only gently as $y \rightarrow 1$, with
$\langle y \rangle \simeq 0.25$ \cite{GQRS96,GQRS98}. Finally, higgsino--like
neutralinos have a rapidly falling $y-$distribution, with $\langle y \rangle
\simeq 0.07$ even after including the energy lost in the decay of the heavier
neutralinos and charginos that are produced in this case; note that
$t-$channel scattering predominantly produces the {\em other} higgsino--like
states, whose masses are however very similar to that of the LSP, so that the
energy lost in the decay only amounts to a few \%. If squarks are lighter than
gluinos, most produced squarks decay into a gluino (rather than an LSP) plus a
quark, with the gluino decaying into at least three particles (including an
LSP again). This would lead to significantly larger energy loss, i.e.~$\langle
y \rangle$ would be even larger. This case has yet to be analyzed in detail.
Here one could use the formalism we developed to treat the decay of the
heavier higgsino--like states, although in case of gluinos the decay might
proceed over several intermediate states \cite{bbkt}.

The transport equation describing LSP propagation through the Earth are
discussed in Sec.~3. We first tried the first order iterative solution of
ref.\cite{iter}, but found that it leads to large violations of flux
conservation in our case; we suspect that this is also true for the rather
similar case of $\nu_\tau$ propagation, where the same method was used in
\cite{reno1,rr}. We therefore solved the transport equation by straightforward
numerical integration. We found that the $y-$distribution is crucial for
understanding the effects of propagation. Indeed, the product of cross section
and $\langle y \rangle$ is a much more reliable ``figure of merit'' here than
the cross section itself. In particular, we saw that propagation effects in a
scenario with higgsino--like LSP are much milder than in a scenario with
bino--like LSP, which has similar total cross section $\sigma$ but much larger
$\sigma \langle y \rangle$; in the former case the LSP flux is suppressed only
for $E \gsim 10^8$ GeV, whereas in the latter, the onset of suppression is at
$E \simeq 10^6$ GeV. Of course, the smaller $\langle y \rangle$ of
higgsino--like LSPs also implies that the visible energy released by LSP
scattering in or near a detector is much smaller than for a bino--like LSP of
equal energy. The visible spectrum for higgsino--like LSPs emerging vertically
out of the Earth would nevertheless extend to at least an order of magnitude
higher energies than that of bino--like LSPs, for similar total cross
section.

Note also that the effective $\langle y \rangle$ for $\nu_\tau$ interactions,
including the effect of $\tau$ decays, is about 0.55, similar to the case of
bino--like LSP. Hence propagation effects will deplete the $\nu_\tau$ spectrum
much faster than the spectrum of higgsino--like LSPs. We therefore expect that
UHE higgsino--like LSPs, which have not yet been considered in the literature,
could also yield a viable signal for visible energies roughly between $10^5$
and $10^7$ GeV. However, a detailed investigation of signals is beyond the
scope of this paper. In the analysis of possible signatures for
ultra--relativistic higgsino--like LSPs, the finite flight path of the heavier
higgsino--like states may also have to be taken into account for scenarios
with mass splitting to the LSP below $\sim 2$ GeV.

In this paper we did not discuss the possibility of a wino--like LSP, as in
scenarios with anomaly mediated supersymmetry breaking \cite{book}. In this
case we expect a strongly suppressed $t-$channel neutral current contribution,
since a neutral wino does not couple to $Z$ bosons, but somewhat enhanced
charged current contribution, since the wino is an $SU(2)$ triplet rather than
a doublet. However, the produced chargino may now lose a significant amount of
energy before decaying, since the mass splitting to the LSP if typically only
a few hundred MeV in this case, leading to a lifetime which is much longer
than that of the $\tau$ lepton. The propagation of these charginos would then
have to be described by a second, coupled, transport equation. A wino--like
LSP can also have a sizable $s-$channel contribution to the cross section, if
squarks are not too heavy. In this case left handed, $SU(2)$ doublet, squarks
would be produced predominantly; if lighter than gluinos, these squarks would
decay into charged or neutral winos, i.e.~into the LSP or the lighter
chargino.

In summary, we provided a treatment of the interactions of ultra--relativistic
higgsino-- or bino--like neutralinos with matter, calculating both the total
cross section and the energy lost in these interactions. We also developed a
new method to solve the transport equations describing the passage of these
neutralinos through matter. This will allow improved estimates of the
prospects of future experiments to observe these neutralinos in potentially
realistic supersymmetric scenarios. Such an observation would help to solve
the forty year old puzzle posed by the cosmic rays at ultra--high energies.

\subsection*{Acknowledgments}
We thank Jon Pumplin for correspondence regarding parameterized parton
distribution functions. MD thanks the high energy theory group of the
University of Hawaii at Manoa for hospitality.

\begin{appendix}

\section{Top $+$ stop production}

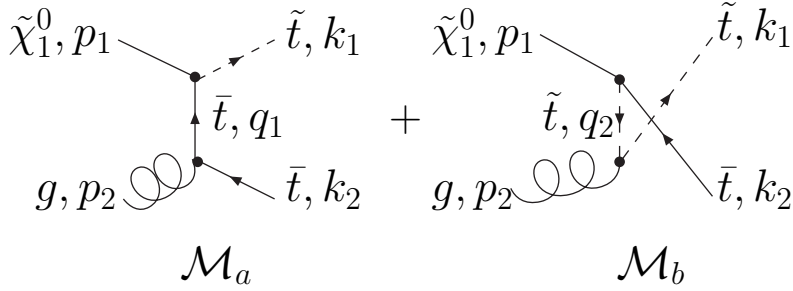
\begin{figure}[h!] 
\begin{center}
\begin{picture}(242,109) (118,-220)
\SetWidth{0.5}
\Vertex(145,-146){2.0}
\ArrowLine(145,-177)(145,-146)
\Line(145,-146)(116,-132)
\Gluon(144,-177)(118,-192){7.5}{2.57}
\ArrowLine(175,-192)(145,-177)
\Vertex(146,-178){2.0}
\DashArrowLine(146,-147)(175,-132){4}
\Text(219,-168)[lb]{\Large{$+$}}
\DashArrowLine(305,-147)(305,-177){4}
\Line(304,-146)(275,-130)
\DashArrowLine(306,-177)(340,-131){4}
\ArrowLine(340,-191)(305,-147)
\Gluon(305,-178)(264,-188){7.5}{2.57}
\Vertex(305,-178){2.0}
\Vertex(305,-147){2.0}
\Text(180,-197)[lb]{\Large{$\bar{t},k_2$}}
\Text(152,-170)[lb]{\Large{$\bar{t},q_1$}}
\Text(343,-136)[lb]{\Large{$\tilde{t},k_1$}}
\Text(277,-170)[lb]{\Large{$\tilde{t},q_2$}}
\Text(235,-139)[lb]{\Large{$\lsp,p_1$}}
\Text(180,-139)[lb]{\Large{$\tilde{t},k_1$}}
\Text(75,-139)[lb]{\Large{$\lsp,p_1$}}
\Text(85,-197)[lb]{\Large{$g,p_2$}}
\Text(235,-197)[lb]{\Large{$g,p_2$}}
\Text(343,-197)[lb]{\Large{$\bar{t},k_2$}}
\Text(305,-225)[lb]{\Large{${\cal M}_{b}$}}
\Text(140,-225)[lb]{\Large{${\cal M}_{a}$}}
\end{picture}
\caption{\label{stopdi}Feynman diagrams contributing to $\lsp + g \rightarrow
  \tilde t + \bar t$. The arrows indicate the flow of baryon number. The
  diagrams for $\lsp + g \rightarrow \bar{\tilde t} + t$ can be obtained by
  reversing the directions of these arrows.}
\end{center}
\end{figure}
The Feynman diagrams for top $+$ stop production via $\lsp-$gluon fusion are
shown in Fig.~\ref{stopdi}.  Here, $p_i$, $k_j$ are four--momenta in the
initial and final state, respectively, and $q_1 = p_2 - k_2$ and $q_2 = p_1 -
k_2$ are the exchanged four--momenta in the two diagrams.  The total squared
partonic production amplitude is then given by $|{\cal M}_a + {\cal M}_b |^2 =
| {\cal M}_a|^2 + | {\cal M}_b|^2 + {\cal M}_a
{\cal M}_b^{\ast} + {\cal M}_a^{\ast} {\cal M}_b$. In the following we list
these three contributions separately.\\

\noindent \fbox{{\bf$| {\cal M}_a|^2$ term:}}\\

\eqa
\frac{1}{4}\sum_{\rm spins}\sum_{\rm colors} | {\cal M}_a |^2 &=&
 -\frac{1}{2} \frac{g_s^2} {\kll t -m_t^2 \klr^2} \cdot \nonumber \\
 && \mbox{} \cdot \biggl[ -\kll |\gei|^2 + |\gzw|^2 \klr \kll k_2 \cdot q_1\:
 p_1 \cdot q_1 - k_2 \cdot p_1 \: q_1 \cdot q_1 + k_2 \cdot q_1\: q_1 \cdot
 p_1 \klr \nonumber \\ 
 && \mbox{} + 2 \gei \gzws \mne m_t q_1^2 + 2\geis \gzw \mne m_t q_1^2
 \nonumber \\ 
 && \mbox{} + 4 \kll |\gei|^2 + |\gzw|^2 \klr m_t^2 \peqe \nonumber \\
 && \mbox{} - 2 \geis \gzw \mne m_t \kzqe - 2 \gei \gzws \mne m_t \kzqe
 \nonumber \\
 && \mbox{} - \kll |\gei|^2 + |\gzw|^2 \klr m_t^2 \pekz \nonumber \\
 && \mbox{} + 2 \geis \gzw \mne m_t^3 + 2 \gei \gzws \mne m_t^3 \biggr]
 \, .  \label{xsgluonteil1}
\eqe

\noindent \fbox{{\bf$| {\cal M}_b|^2$ term:}}\\

\eqa
\frac{1}{4}\sum_{\rm spins}\sum_{\rm colors} | {\cal M}_b |^2
&=& -\frac {1}{4} \frac {g_s^2} {\kll u-\mtt^2 \klr^2}
 \kll t + 2 \kzqz + \mtt^2 \klr \nonumber \\ 
&& \cdot \bigl{[} \kll| \gei|^2+ |\gzw|^2\klr   \peke
+ 2 \gei \gzws \mne m_t \bigr{]} \, .
\label{xsgluonteil2}
\eqe
\fbox{{\bf${\cal M}_a {\cal M}_b^{\ast} + {\cal M}_a^{\ast} {\cal M}_b$
    terms:}}\\
\eqa
\frac{1}{4}\sum_{\rm spins}\sum_{\rm colors }
\kll  {\cal M}_a {\cal M}_b^* + {\cal M}_a^* {\cal M}_b  \klr
&=&-\frac{1}{2}\frac{g_s^2}{(t-m_t^2)(u-\mtt^2)}\Bigl[
\kll|\gei|^2+|\gzw|^2\klr    \nonumber \\
 &&\cdot  \biggl( \pekz\qeqz+\peqe\kekz-\peke\kzqe+\pekz\keqe\nonumber \\
 &&\mbox{}-\peqz\frac{m_t^2}{2}-\pekz\frac{m_t^2}{2}\Bigr)  \nonumber \\
 &&\mbox{}+  \kll\qeqz+\keqe-\kzqz-\kekz\klr  \kll m_t\mne\gei\gzws \klr
 \biggr]  \, . \label{xsgluonteil3}
\eqe
Here, $g_s$ is the strong (QCD) coupling constant, $t=q_1^2$, $u=q_2^2$, $m_t$
is the mass of the top quark and $\mtt$ that of the produced stop squark. The
couplings $G_i^{L,R}$ depend on whether the lighter ($i=1$) or heavier ($i=2$)
stop squark is to be produced:
\eqa \label{stcoup}
G_1^L &=& - \sqrt{2} g_2 \kll \frac {1} {2} N_{12}^{\ast} + \frac {1}
{6} \tan\theta_W N_{11}^{\ast} \klr \cos\theta_{\tilde t}
       - \frac {g_2 m_t} {\sqrt{2} M_W \sin\beta}  N_{14}^{\ast} \sin
       \theta_{\tilde t}\, , \nonumber \\
G_2^L &=& \sqrt{2} g_2 \kll \frac {1} {2} N_{12}^{\ast} + \frac {1}
{6} \tan\theta_W N_{11}^{\ast} \klr \sin\theta_{\tilde t}
       - \frac {g_2 m_t} {\sqrt{2} M_W \sin\beta}  N_{14}^{\ast} \cos
       \theta_{\tilde t}\, , \nonumber \\
G_1^R &=& \frac {2\sqrt{2}} {3} g_2 \tan\theta_W N_{11}
\sin \theta_{\tilde t} - \frac{g_2 m_t} {\sqrt{2} M_W
  \sin\beta} N_{14} \cos\theta_{\tilde t}\, , \nonumber \\
G_1^R &=&  \frac {2\sqrt{2}} {3} g_2 \tan\theta_W N_{11}
\cos \theta_{\tilde t} + \frac{g_2 m_t} {\sqrt{2} M_W
  \sin\beta} N_{14} \sin\theta_{\tilde t}\, , 
\eqe
where $\theta_{\tilde t}$ is the stop mixing angle (i.e.~$\tilde t_1 = \cos
\theta_{\tilde t} \tilde t_L + \sin \theta_{\tilde t} \tilde t_R$).

\vspace*{8mm}
\begin{figure}[h!] 
\begin{center}
\includegraphics[width=8cm]{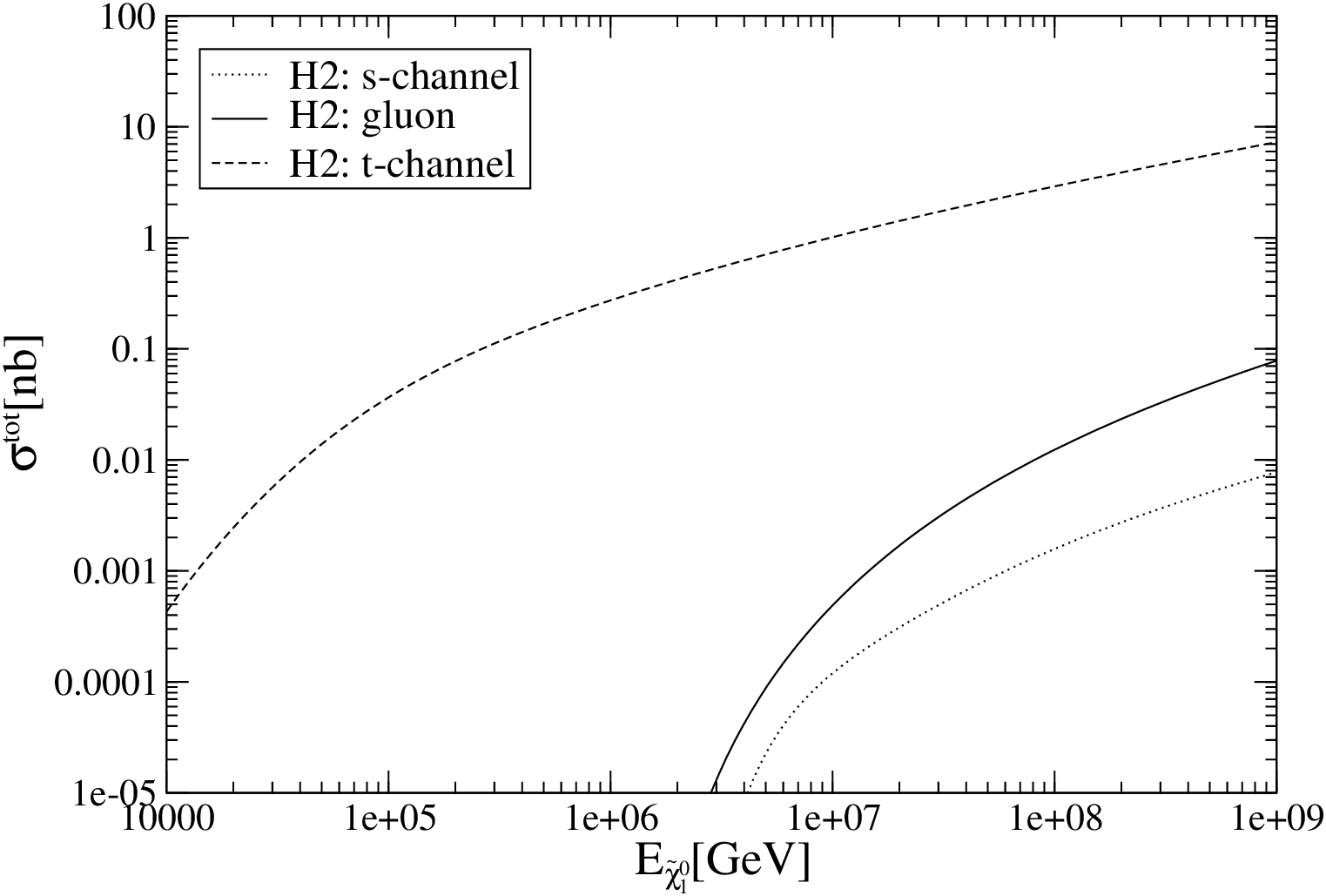}
\includegraphics[width=8cm]{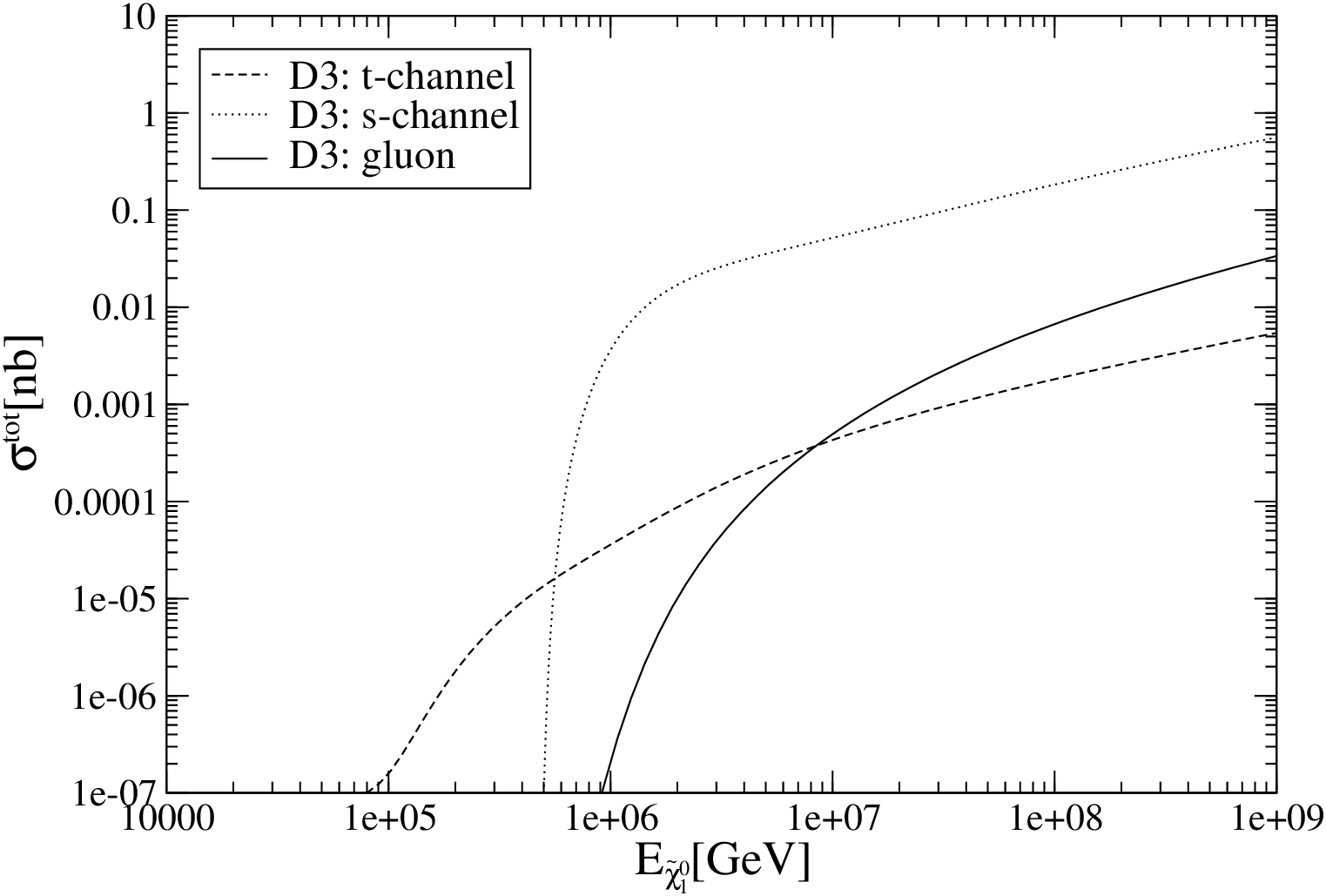}
\caption{\label{figapp}The total cross section for $\lsp+ g \rightarrow \tilde
  t \bar t, \ \bar{\tilde t} t$ (solid curves) in comparison with the $s-$
  (dotted) and $t-$channel (dashed curves) contributions.}
\end{center}
\end{figure}

The added cross section of $\lsp g\rightarrow\tilde{t}\bar{t}$ and $\lsp
g\rightarrow\bar{\tilde{t}}t$ for a higgsino-- and bino--like neutralino are
displayed in the left and right frames of Fig.~\ref{figapp}, respectively; we
used $m_{\tilde t_1} = 1.2$ TeV, $m_{\tilde t_2} = 1.7$ TeV, $\theta_{\tilde
  t} = 0.03$ in scenario H2, and $m_{\tilde t_1} = 0.76$ TeV, $m_{\tilde t_2}
= 0.91$ TeV, $\theta_{\tilde t} = 1.4$ in scenario D3. We see that in both
cases $t + \tilde t$ production only make sub--dominant contributions to the
total cross section, although in scenario H2 it dominates the total squark
production cross section; the reason is that higgsinos couple proportional to
the mass of the quark, as shown by the second term of each of
Eqs.(\ref{stcoup}). Note also that the average energy loss in $t + \tilde t$
production will be even larger than in $s-$channel reactions involving light
quarks. These processes can therefore be important for higgsino--like LSP if
$2 m_{\tilde t_1} \lsim |\mu| < |M_1|, \, |M_2|$ (which does not happen in
mSUGRA \cite{book}), or in scenarios with bino--like LSP if $\tilde t_1$ is
much lighter than first or second generation squarks (which can happen even in
mSUGRA, if $|A_0|$ is large).

\end{appendix}

\end{document}